\documentclass[fleqn,usenatbib]{mnras}
\usepackage[T1]{fontenc}
\usepackage{ae,aecompl}
\usepackage{bm}
\usepackage{graphicx}

\usepackage{amsmath}	% Advanced maths commands
\usepackage{amssymb}	% Extra maths symbols
\usepackage{txfonts}

\newcommand{\SL}[1]{\textcolor{blue}{#1}}

\title[Orbital stability of two circumbinary planets]{Orbital stability of two circumbinary planets around misaligned eccentric binaries}

\author[Chen et al.]{Cheng Chen$^{1,2}$\thanks{Email: c.chen6@leeds.ac.uk}, Stephen H. Lubow$^3$,  Rebecca G. Martin$^1$ and C. J. Nixon$^2$ 
\\ $^1$Department of Physics and Astronomy,  University of Nevada, Las Vegas, 4505 South Maryland Parkway, Las Vegas, NV 89154, USA 
\\ $^{2}$School of Physics and Astronomy, University of Leeds, Leeds LS2 9JT, UK
\\ $^{3}$Space Telescope Science Institute, 3700 San Martin Drive, Baltimore, MD 21218, USA\\
}

\date{Accepted XXX. Received YYY; in original form ZZZ}

\pubyear{2022}

\begin{document}
\label{firstpage}
\pagerange{\pageref{firstpage}--\pageref{lastpage}} 
\maketitle


\begin{abstract}
With $n$-body simulations we investigate the stability of tilted circumbinary planetary systems consisting of two nonzero mass planets. The planets are initially in circular orbits that are coplanar to each other, as would be expected if they form in a flat but tilted circumbinary gas disc and decouple from the disc within a time difference that is much less than the disc nodal precession period. We constrain the parameters of stable multiple planet circumbinary systems. Both planet-planet and planet-binary interactions can cause complex planet tilt oscillations which can destabilise the orbits of one or both planets. The system is considerably more unstable than the effects of these individual interactions would suggest, due to the interplay between these two interactions. The stability of the system is sensitive to the binary eccentricity, the orbital tilt and the semi-major axes of the two circumbinary planets. With an inner planet semi-major axis of $5\,a_{\rm b}$, where $a_{\rm b}$ is semi-major axis of the binary, the system is generally stable if the outer planet is located at  $\gtrsim 8\,a_{\rm b}$, beyond the 2:1 mean motion resonance with the inner planet. For larger  inner planet semi-major axis the system is less stable because the von-Zeipel--Kozai--Lidov mechanism plays a significant role, particularly for low binary-eccentricity cases.  For the unstable cases, the most likely outcome is that one planet is ejected and the other remains bound  on a highly eccentric orbit. Therefore we suggest that this instability is an efficient mechanism for producing free-floating planets.  
\end{abstract}
%TESS and PLATO  may find misaligned circumbinary planetary systems in the near future.  


\section{Introduction}

% 1st paragraph - motivation for studying misaligned CB planets
Giant planets around a binary star may form with the same initial orbital properties as the circumbinary disc from which they form. Recent observations show that there are many misaligned circumbinary discs  \citep[e.g.,][]{Chiang2004,Winn2004,Capelo2012,Kennedy2012,Brinch2016,Kennedy2019,Zhu2022,Kenworthy2022}. 
Although about 68\% of short period binaries (period $<20\,\rm days $) have aligned disks (within 3$^{\circ}$), those with longer orbital periods have a larger range of inclinations and binary eccentricities \citep{Czekala2019}. The formation of misaligned discs may be due to chaotic accretion  \citep{clarke1993,Bate2003,Bate2018} or stellar flybys \citep{Cuello2019b,Nealon2020}. Protoplanetary discs typically nodally precess as a solid body \citep[e.g.][]{PT1995,LO2000}. Dissipation in the disc leads to tilt evolution \citep[e.g.][]{Nixonetal2011a,MartinandLubow2017} but for a sufficiently extended disc, the lifetime may be longer than the alignment timescale. Thus, circumbinary planets (CBPs) may form in misaligned discs.

Although CBPs with a wide range of inclinations are expected to be in binaries with longer orbital periods, they are harder to detect by the transit method because the transit probability is smaller and the planet orbital period is longer \citep{MartinD2014,MartinD2019}.  To date, all the CBPs which have been detected are nearly coplanar to the binary orbit due to the small orbital period of the Kepler binaries \citep{Czekala2019}. Eclipse timing variations of the binary may be a better method to to distinguish polar planets from coplanar planets \citep{Zhang2019,MartinD2021}. In the current observations, the Kepler-47 system has multiple CBPs but the binary orbit is nearly circular ($e < 0.03$) and the three Neptune-size planets are nearly coplanar \citep{Orosz2012a, Orosz2012b, Kostov2013}. Moreover, the TOI-1338 system has two saturnian CBPs detected by transit and radial velocity methods \citep{Kostov2020, Standing2023}, the NN Ser binary system that is comprised of a red dwarf and white dwarf hosts two Jupiter mass CBPs \citep{Mustill2013} and Kepler-451 has three Jupiter mass CBPs  \citep{Baran2015, Esmer2022}. All of these planets are nearly coplanar to the binary orbital plane.

% 3rd  paragraph - describe test particle cases

For a misaligned (massless) test  particle orbiting around a circular orbit binary, its angular momentum vector precesses around the binary angular momentum vector. The nodal precession is either prograde or retrograde depending upon the initial particle inclination. These are {\it circulating} orbits. The longitude of the ascending node fully circulates over 360$^\circ$ during the nodal precession. For a binary with nonzero eccentricity, a test particle orbit  with a sufficiently large initial inclination may undergo nodal libration where the angular momentum vector of the test particle precesses about the binary eccentricity vector.  During this process, the orbit undergoes tilt oscillations while the longitude of the ascending node is limited in a range of angles less than 360$^\circ$ \citep{Verrier2009,Farago2010,Doolin2011,Naoz2017,deelia2019}. These are {\it librating} orbits. The minimum inclination (critical inclination) required for libration decreases with increasing binary eccentricity. Therefore, a test particle orbit with even a small inclination can librate around a highly eccentric binary. 
 
% 4th paragraph - descrie the effect of mass on the one planet case

The dynamics of a CBP around an eccentric binary are somewhat affected by the mass of the planet \citep{Chen20192}.  For a misaligned non-zero mass planet orbiting around a eccentric orbit binary, the critical angle for the planet to librate depends on the binary eccentricity and angular momentum ratio of the planet to the binary. The angle of the stationary inclination, or polar alignment, occurs at less than 90$^{\circ}$ if the planet is massive \citep{Farago2010, Lubow2018, Zanazzi2018,MartinandLubow2019}.

Generally, a single circumbinary planet  on an initially circular orbit is stable if its initial orbital radius is greater than 5 times the binary semi-major axis \citep{Doolin2011, Chen20201}. Stable orbits that are closest to the binary  are nearly retrograde and circulating for small binary eccentricity \citep{Cuello2019, Giuppone2019}. On the other hand, the most stable orbits are highly inclined, near the polar inclination, for high binary eccentricity \citep{Chen20201}. 

With two CBPs around an eccentric orbit binary, the system is more complex because the two planets interact with not only the binary but also each other. Both of these interactions can cause complex tilt oscillations of the planets \citep{Chen2022}. Planet-planet interactions may result in a planet being ejected from system, as has been seen already in coplanar CBP systems  \citep[e.g.,][]{Smullen2016, Sutherland2016, Gong20171, Gong20172}. At least one planet is ejected in 87\% of multi-planet circumbinary systems for low-mass and short period-binaries \citep{Fleming2018}. Nevertheless, the existence of multiple CPBs around Kepler-47, TOI-1338, NN Ser and Kepler-451 suggests that multi-planet circumbinary systems may not be rare and the Kepler data show that about half of planets are known to have siblings \citep[e.g.][]{Berger2018, Thompson2018}. We expect that more circumbinary planetary systems will be found in the future. Hence, a comprehensive study of such systems is necessary for understanding the orbital dynamics and evolution of two or even more circumbinary planets.

In this study, we investigate the orbital stability of two misaligned CBPs with initially circular orbits around circular or eccentric binaries. The two CBPs begin coplanar to each other but misaligned to the binary orbit \citep[e.g.][]{Chen2022}. 
We first describe the set up of the four-body simulations in Section~\ref{sta}. We describe our results and stability maps with different initial semi-major axes of the two planets in Section~\ref{sim}.  We consider the final orbital distributions of the surviving and stable planets in Section~\ref{sing}. Finally, we present our discussion and conclusions in Section~\ref{diss}.

\label{int}

\section{Simulation set--up and parameter space explored}
\label{sta}

\begin{table}
\centering
\caption{Parameters of the simulations. The first column contains the name of the Model, the second and third columns indicate the binary eccentricity and initial semi-major axis of the inner planet. The fourth and fifth columns represent the minimum and maximum separations of the outer planet that we consider with an interval of $0.1 \,a_{\rm b}$.}
\begin{tabular}{cccccc} 
\hline
\textbf{Model} & $e_{\rm b}$ & $a_{\rm p1}$ ($a_{\rm b}$) & Minimum $a_{\rm p2}$ ($a_{\rm b}$) & Maximum $a_{\rm p2}$ ($a_{\rm b}$)  \\
\hline
\hline
X1 &  0.0  & 5.0 & 5.9 & 10.0 \\
X2 &  0.2  & 5.0 & 5.9 & 10.0 \\
X3 &  0.5  & 5.0 & 5.9 & 10.0  \\
X4 &  0.8  & 5.0 & 5.9 & 10.0  \\
\hline
Y1 &  0.0  & 10.0 & 12.0 & 20.0  \\
Y2 &  0.2  & 10.0 & 12.0 & 20.0 \\
Y3 &  0.5  & 10.0 & 12.0 & 20.0  \\
Y4 &  0.8  & 10.9 & 12.0 & 20.0  \\
\hline
Z1 &  0.0  & 20.0 &  25.0 & 40.0  \\
Z2 &  0.2  & 20.0 &  25.0 & 40.0  \\
Z3 &  0.5  & 20.0 &  25.0 & 40.0 \\
Z4 &  0.8  & 20.0 &  25.0 & 40.0  \\
\hline
\label{table1}
\end{tabular}
\end{table}

In this section we first describe the simulation set--up and the parameter space that we explore. To study the orbital stability of two planets orbiting around a circular or eccentric binary star system, we use a {\sc whfast} integrator which is a second order symplectic Wisdom Holman integrator with 11th order symplectic correctors in the ${\sc n}$-body simulation package, {\sc rebound} \citep{Rein2015b}. We set the timestep to be 0.7$\%$ of the initial orbital period of the inner planet.

We solve the gravitational equations of  four bodies in the frame of the centre of mass of the four-body system for which the central binary has components of equal mass $m_1$ and $m_2$ with total mass $m_{\rm b}=m_{\rm 1}+m_{\rm 2}$. The semimajor axis of the binary is $a_{\rm b}$, the eccentricity of the binary is $e_{\rm b}$ and the orbital period of the binary is $T_{\rm b}$. Our simulations are scale free such that all masses are scaled to the total binary mass, $m_{\rm b}$, and all distances are scaled to the binary semi-major axis, $a_{\rm b}$.

The orbital elements of the two planets are determined by assuming that they lie on Keplerian orbits around the centre of mass of the binary. The planets have equal masses $m_{\rm p1} = m_{\rm p2} = 0.001\,m_{\rm b}$. Our simulations do not account for collisions or the formation of S-type planets (planets that orbit around one star of a binary). Their orbits are defined by six orbital elements: the semi-major axes $a_{\rm p1}$ and  $a_{\rm p2}$, inclinations $i_{\rm p1}$ and $i_{\rm p2}$ relative to the binary orbital plane, eccentricities $e_{\rm p1}$ and $e_{\rm p2}$, longitude of the ascending nodes $\phi_{\rm p1}$ and $\phi_{\rm p2}$ measured from the binary semi--major axis, argument of periapsides $\omega_{\rm p1}$ and  $\omega_{\rm p2}$, and true anomalies $\nu_{\rm p1}$ and $\nu_{\rm p2}$. The orbits of the two planets are initially coplanar to each other and circular so initially $e_{\rm p}=0$, $\omega_{\rm p}=0$ and $\nu_{\rm p}=0$. 

As planets form within the disc, they are initially coupled to the disc by gravitational forces and precess with it. Once they open a gap and decouple from the disc, they can precess at a rate that differs from the disc's precession rate. If they decouple within a disc nodal precession period of each other, they will begin their orbital evolution in a mutually coplanar state. However, the effects of the disc may still play a role in their orbital evolution \citep{Lubow2016}.

We consider three initial values of $a_{\rm p1}= 5\,a_{\rm b}$, $10\,a_{\rm b}$, and $20\,a_{\rm b}$ and three initial ranges of $a_{\rm p2}$  that depend on the initial value of $a_{\rm p1}$ (see Table 1) with the interval 0.1$a_{\rm b}$. We vary the initial values of  the tilt $i_{\rm p1}=i_{\rm p2}$ from $0^{\circ}$ to $180^{\circ}$ with an interval of 2.5$^{\circ}$. The binary orbit is not fixed since the binary feels the gravity of massive planets.

In order to analyse the orbital motion of the planet, we work in a frame defined by the instantaneous eccentricity and angular momentum vectors of the binary ($\bm{e_{\rm b}}$ and $\bm{l_{\rm b}}$).  
The frame  has the three axes along unit vectors
$\bm{\hat{e}_{\rm b}}$, $\bm{\hat{l}_{\rm b}}\times \bm{\hat{e}_{\rm b}}$, and $\bm{\hat{l}_{\rm b}}$. For
the planet angular momentum $\bm{l_{\rm p}}$,  the inclination of the planet's orbital plane relative to the binary orbital plane is given by
\begin{equation}
i_{\rm p} =\cos^{-1}(\bm{\hat l_{\rm b}}\cdot \bm{\hat l_{\rm p}}),
\end{equation}
where $\bm{\hat l_{\rm b}}$ is a unit vector in the direction of the
angular momentum of the binary and $\bm{\hat l_{\rm p}}$ is a unit
vector in the direction of the angular momentum of the
planet.
The mutual inclination between the two planets   is given by
\begin{equation}
\Delta i_{\rm p} =\cos^{-1}(\bm{\hat l_{\rm p1}}\cdot \bm{\hat l_{\rm p2}}), \label{Deltaip}
\end{equation}
where $\bm{\hat l_{\rm pi}}$ for $i=1,2$ is a unit vector  in the direction of the angular momentum of each
planet.
The inclination of the binary relative to the total angular momentum $\bm{l}$ is
\begin{equation}
i_{\rm b} =\cos^{-1}(\bm{\hat l}\cdot \bm{\hat l_{\rm b}}),
\end{equation}
where $\bm{\hat l}$ is a unit
vector in the direction of the total angular momentum ($\bm{l}=\bm{l}_{\rm b}+\bm{l}_{\rm p}$). Similarly, the phase angle (longitude of ascending node) of the planet in the
same reference frame is given by
\begin{equation}
\phi_{\rm p}=\tan^{-1}\left(\frac{ \bm{\hat{l}}_{\rm p}\cdot (\bm{\hat{l}}_{\rm b}\times \bm{\hat{e}}_{\rm b})}{ \bm{\hat{l}}_{\rm p}\cdot \bm{\hat{e}}_{\rm b}}  \right) + 90^{\circ}.
\end{equation}

\section{Simulation results}
\label{sim}
In this section, we investigate the orbital stability of two planets around a single star and two planets around a binary. We are interested in the destabilising effects of the planet-planet interactions that involve the central binary. For comparison, we first look at the case of two planets orbiting a single star in Sec.~\ref{single}. The criterion for the long term stability of two circular and coplanar planets around a single star is that their separation satisfies 
\begin{equation}
\Delta= a_{\rm p2}-a_{\rm p1} \geq \Delta_{\rm crit} 
\label{Delta}
\end{equation}
with
\begin{equation}
\Delta_{\rm crit}=2\sqrt{3} R_{\rm Hill}
\label{deltacrit}
\end{equation}
\citep{Marchal1982,Gladman1993,Chambers1996}, where  the mutual Hill sphere radius is given by
\begin{equation}
R_{\rm Hill} = \left(\frac{m_{\rm p1}+m_{\rm p2}}{3 \,m_{\rm b}}\right)^{1/3}\left(\frac{a_{\rm p1}+a_{\rm p2}}{2} \right).
\end{equation}
The stability criterion in equation~(\ref{Delta}) has been confirmed numerically \citep[e.g.][]{Chambers1996,Marzari2016}, but has not been specifically tested for two planet systems around a binary.

For the single CBP case, numerical three-body simulations show that a single CBP is stable if it has semi-major axis $a_{\rm p} \gtrsim 5\ a_{\rm b}$ around an equal mass binary system  for all  initial $i_{\rm p}$ and $e_{\rm b}$ \citep{Chen20201}. Hence, for the  binary case  we then consider inner planets  at $a_{\rm p1} = 5$, $10$ and $20 \,a_{\rm b}$ after Sec.~\ref{single}. We classify the orbit of  a planet to be unstable if the eccentricity of the planet, $e_{\rm p}$, is larger than 0.99 or if the semi-major axis is less than $\,a_{\rm b}$ or larger than 1000$\,a_{\rm b}$. Once a planet meets one of these criteria, we consider this four-body system to be unstable.

\subsection{Single star system with two planets}
\label{single}
To isolate the effects of the binary, we consider a stability map of two planets around a single star with the mass $\rm M_{\odot}$.  The planets both have a mass of $0.001 \rm M_{\odot}$. The inner planet is initially at $a_{\rm p1}$ =10 au while we consider orbits of the outer planet for which the initial semi–major axis is in the range 12 au $\le$ $a_{\rm p2}$ $\le$ 20 au. (In the code, we use dimensionless values of length and mass. We take the unit of distance to represent the dimensional value of 1 au and the unit of mass to represent the mass of the central star that is dimensionally $ 1 \rm M_{\odot}$.) We make an arbitrary choice for the reference plane that is defined for which the inclination is zero. The planets are initially coplanar to each other. We consider a range of initial planet inclination, $0^\circ \le i_{\rm p}\le 180^\circ$. While there is no physical difference between different inclinations around a single star, this setup allows us to test how well the outcomes are determined and allows us to check the stability criteria given by equation~(\ref{Delta}) that was determined for planets around a single star. Finally, it is useful for comparison to the binary cases that we consider later.

\begin{figure}
  \centering
    \includegraphics[width=8.cm]{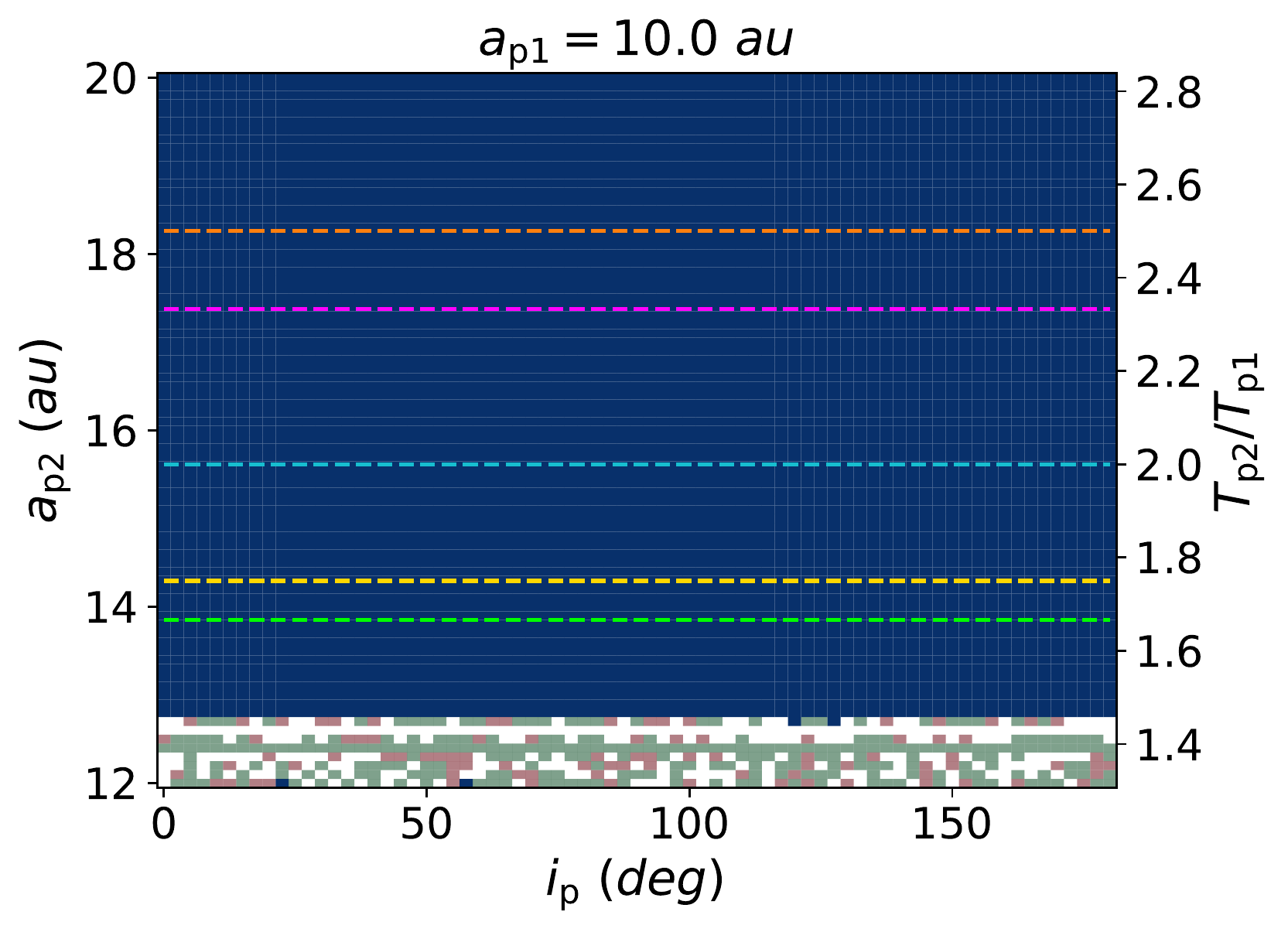}
    \caption{ Stability of two planets of mass $0.001\,\rm M_\odot$ orbiting a single star of mass $1\,\rm M_\odot$. The inner planet has a semi-major axis of $10\,\rm au$.
    The left $y$-axis is the initial semi-major axis of the outer planet while the right $y$-axis is the orbital period ratio of the inner planet ($T_2$) to the outer planet ($T_1$). The $x$-axis is the initial inclination of planets. Blue pixels represent stable orbits for both planets, green pixels represent stable orbits for the inner planet and red pixels represent the stable orbits for the outer planet. White pixels represent unstable orbits for both planets. The five horizontal dashed lines indicate the 5:2, 7:3, 2:1, 7:4 and 5:3 mean motion resonances between the planets in orange, magenta, cyan, yellow and green, respectively.
 }
     \label{fig:single}
\end{figure}

In Fig.~\ref{fig:single}, the different pixel colours represent the final orbital status of the two planets. Blue pixels indicate that both planets are stable at the end of the simulation and white pixels represent two unstable planets. Green pixels indicate that only the inner planet survives at the end of the simulation and red pixels indicate that only the outer planet survives at the end of the simulation. 

After we integrate for 14 million years, the two planets are quite stable except in the region $a_{\rm p2} < 13$ au. The instability in this region  is due to planet-planet interactions that involves Hill instability and chaotic behaviour \citep{Chambers1996}.  For our parameters, equation~(\ref{Delta}) predicts stability for $a_{\rm p2}> 13.6$ au and therefore our numerical results concur with this criterion.

 Due to spherical symmetry, these results around a single star should be independent of the initial inclination, $i_{\rm p}$, of the two planets. The range of unstable orbital radii for the outer planet is independent of $i_{\rm p}$, reflecting the spherical symmetry. On the other hand, within the unstable region,  the plot shows a mixture of unstable outcomes involving two planets being ejected, only the inner planet being ejected and only the outer planet being ejected.  These results do not reflect the spherical symmetry. The reason for the change in behaviour is simply due to round off error in the initial conditions at different initial inclinations. This effect shows that we cannot determine the details of a specific unstable outcome with any confidence.  This is likely to be a generic issue for all instability plots involving systems like this.

\subsection{Binary system with two planets: $a_{\rm p1}$ = 5$\,a_{\rm b}$}
\begin{figure*}
  \centering
    \includegraphics[width=8.7cm]{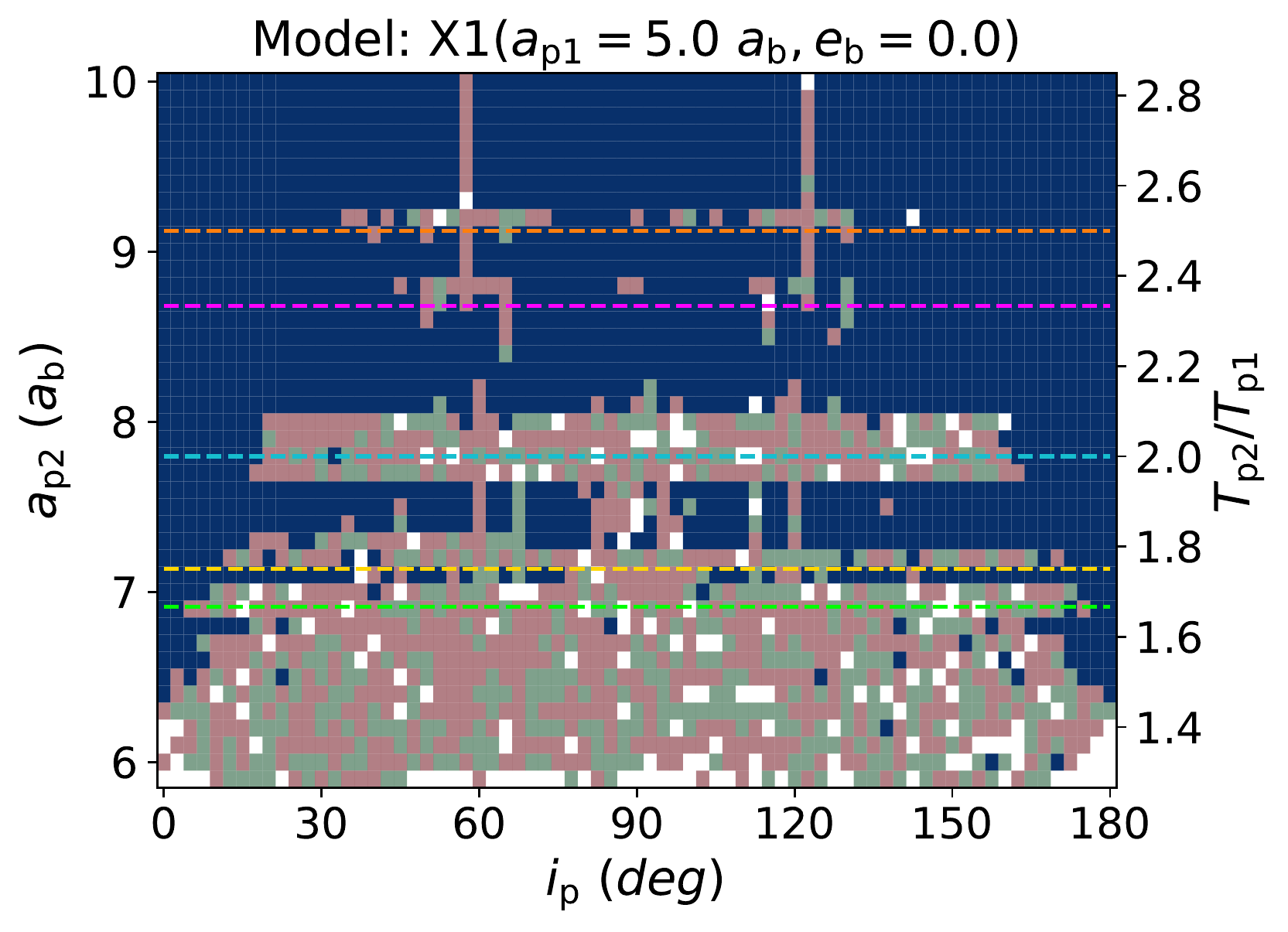}
    \includegraphics[width=8.7cm]{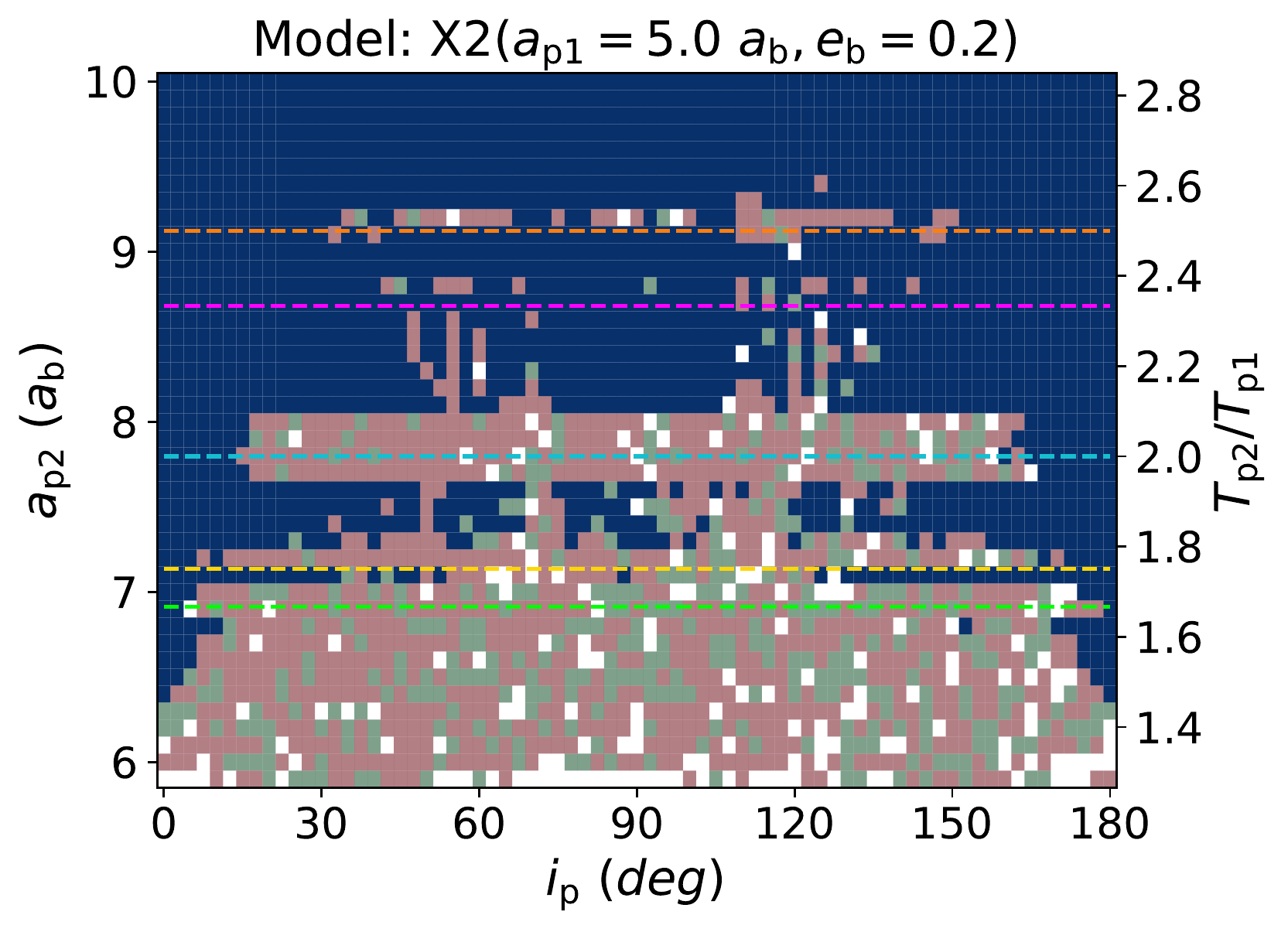}
    \includegraphics[width=8.7cm]{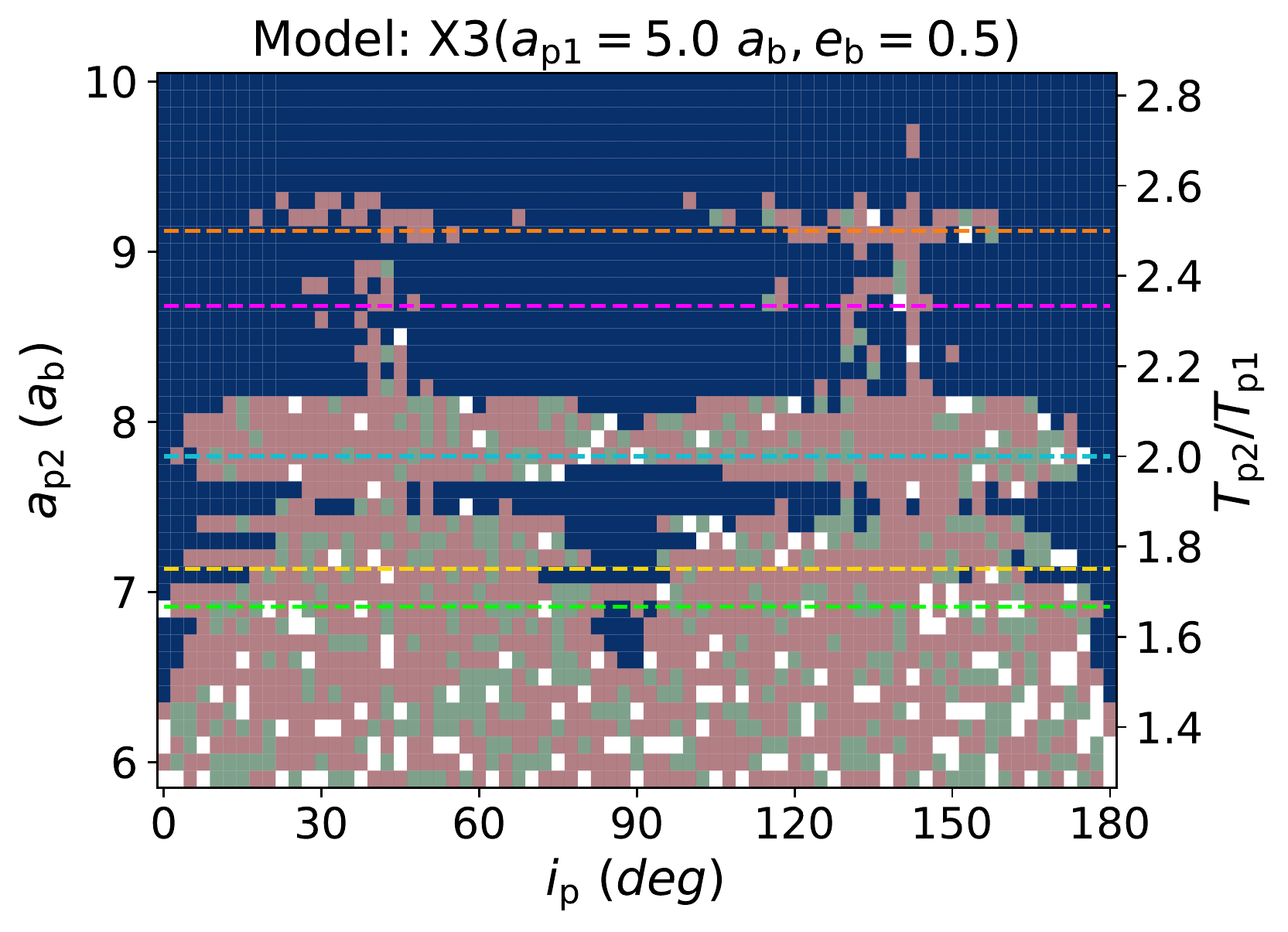}       
    \includegraphics[width=8.7cm]{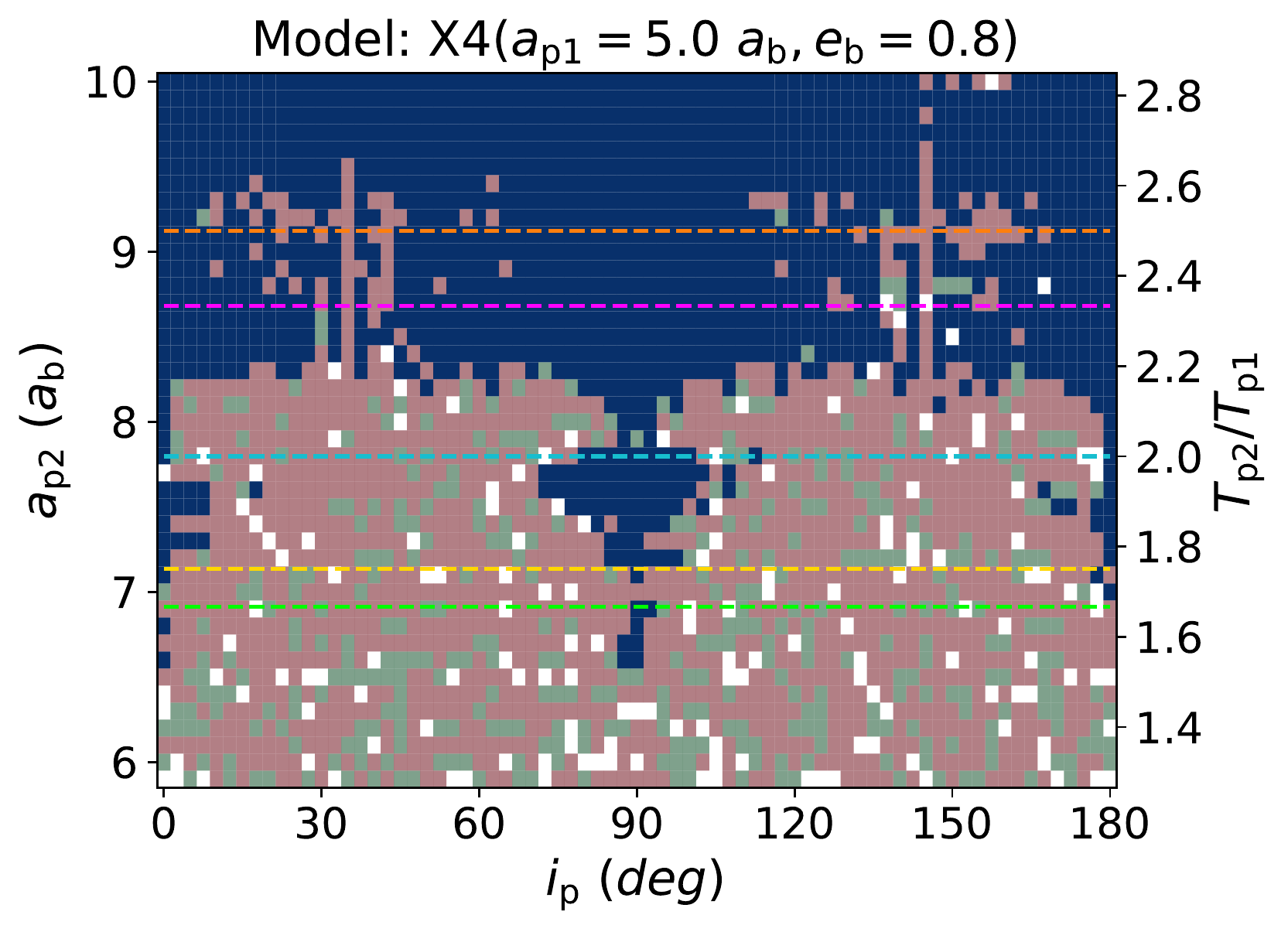}
    \caption{Stability maps of two circumbinary planets in a binary system with $e_{\rm b} =0.0$ (upper-left), 0.2 (upper-right panel), 0.5 (lower-left) and 0.8 (lower-right). The inner planet has initial semi-major axis $a_{\rm p1} = 5\, a_{\rm b}$. The five horizontal dashed lines and the pixels colours are the same as Fig.~\ref{fig:single}}.
     \label{fig:mapa5}
\end{figure*}

We now take the effect of a binary with total mass $m_{\rm b}$  into account and first consider systems in which the inner planet is initially placed at $a_{\rm p1}=5\,a_{\rm b}$. The outer planet is taken to have an initial semi--major axis in the range $5.9\, a_{\rm b} \le a \le 10.0\, a_{\rm b}$. We integrate the orbits for a total time of $5\times 10^6\, T_{\rm b}$, where $T_{\rm b}$ is the orbital period of the binary. The stability maps in Fig.~\ref{fig:mapa5} show models X1--X4, with different binary eccentricities $e_{\rm b}$ = 0.0 (upper-left), 0.2 (upper-right), 0.5 (lower-left), and 0.8 (lower-right). 

The right vertical axis of each panel in Fig.~\ref{fig:mapa5} shows the orbital period ratio of the inner planet to the outer planet, $T_{\rm p2} / T_{\rm p1}$. From the top to bottom, there are several resonance regions that indicate instability. The five horizontal lines show the 5:2, 7:3, 2:1, 7:4 and 5:3 mean motion resonances between the planets in colours of orange, magenta, cyan, yellow and green, respectively. The widest resonance region is the 2:1. If the outer planet is closer in than the 2:1 resonance, $\lesssim 8\,a_{\rm b}$, then most of the cases are unstable, \SL{mainly for nonclopanar cases.} This instability results from the closer gravitational interactions and may involve stronger tilt oscillations between the two planets. 

 In the cases with a small binary eccentricity of $e_{\rm b}=0$ and 0.2, only the systems that are close to coplanar or close to retrograde coplanar are stable for all plotted separations with $a_{\rm p2}>a_{\rm p1}+\Delta_{\rm crit}=6.7\,a_{\rm b}$ using equation~(\ref{Delta}). Their greater stability is due to the relatively small variations of the inclination difference between the planets.

Between the 2:1 and 7:4 resonance regions, the number of unstable orbits increases with increasing $e_{\rm b}$. This is  likely because orbital libration can result in large variations in $i_{\rm p}$ if $e_{\rm b}$ is large \citep[see the phase plots in][]{Chen20192}. The inclination variation is small when $i_{\rm p}$ is close to a stationary inclination. This is why there are more stable orbits near the polar region ($i_{\rm p}\approx 90^\circ$) in the $e_{\rm b}$ = 0.5 and 0.8 cases. The polar region is stable down to near $a_{\rm p2}=6.5 \,a_{\rm b}$.  However, the 2:1 resonance unstable region is wider outside of the the polar region.

The binary eccentricity has another effect on stability. Outside of the 2:1 resonance, in $a_{\rm p2}\gtrsim 8\,a_{\rm b}$, there are two obvious unstable belts located around $i_{\rm p}=60^\circ$ (and 120$^\circ$), 45$^\circ$ (and 135$^\circ$), and 30$^\circ$ (and 150$^\circ$) in models X2, X3 and X4, respectively. Those angles are are close to the critical minimum inclination angle between the circulating and librating orbits \citep[see section 3.2 in][]{Chen20192}.
Unlike the smooth transition in the single CBP stability maps in \citet{Chen20201}, CBPs near these critical inclinations can go unstable because the two planets can affect each other. Specifically, in Fig.~12 in \citet{Chen2022}, we found that the outer planet in the system has both a lower critical inclination for libration and a lower polar aligned inclination than the inner planet. The two planets can therefore be undergoing different types of nodal oscillation (circulation or libration) with the binary and they can be destabilised in their orbits. In most cases, the inner planets go unstable while the outer planets remain stable. 

We notice that there are also two narrow belts in model X1 when the two planets have a certain $i_{\rm p}$. We run several test simulations and we find that $e_{\rm p1}$ gets excited within a short time (> 500 $T_{\rm b}$). The reason is unclear and we should investigate this kind of orbital evolution in the future which may be a dynamical effect we do not consider in our model. 

Although we cannot determine the details of a specific unstable outcome with any confidence, we would like to know if the ratio of different outcomes varies with the initial $a_{\rm p2}$. Thus, we added a slight displacement (0.001$a_{\rm b}$) to the initial $a_{\rm p2}$ values and reran the simulations. We found that the ratio of the cases in which only the inner planet survived to those in which only the outer planet survived (here after $R_{\rm i/o}$) does not vary significantly even though the unstable outcome for a specific  set of initial conditions may not be the same as the original models. Moreover, the ratio generally decreases with increasing $e_{\rm b}$ which means that the outer planet has a higher chance of survival than the inner planet in a high $e_{\rm b}$ system. These ratios are approximately  $R_{\rm i/o}=1/2$ (models X1 and X2), $1/3$ (model X3) and $1/4$ (model X4).

\subsection{Binary system with two planets: $a_{\rm p1}$ = 10$\,a_{\rm b}$}

\begin{figure*}
  \centering
    \includegraphics[width=8.7cm]{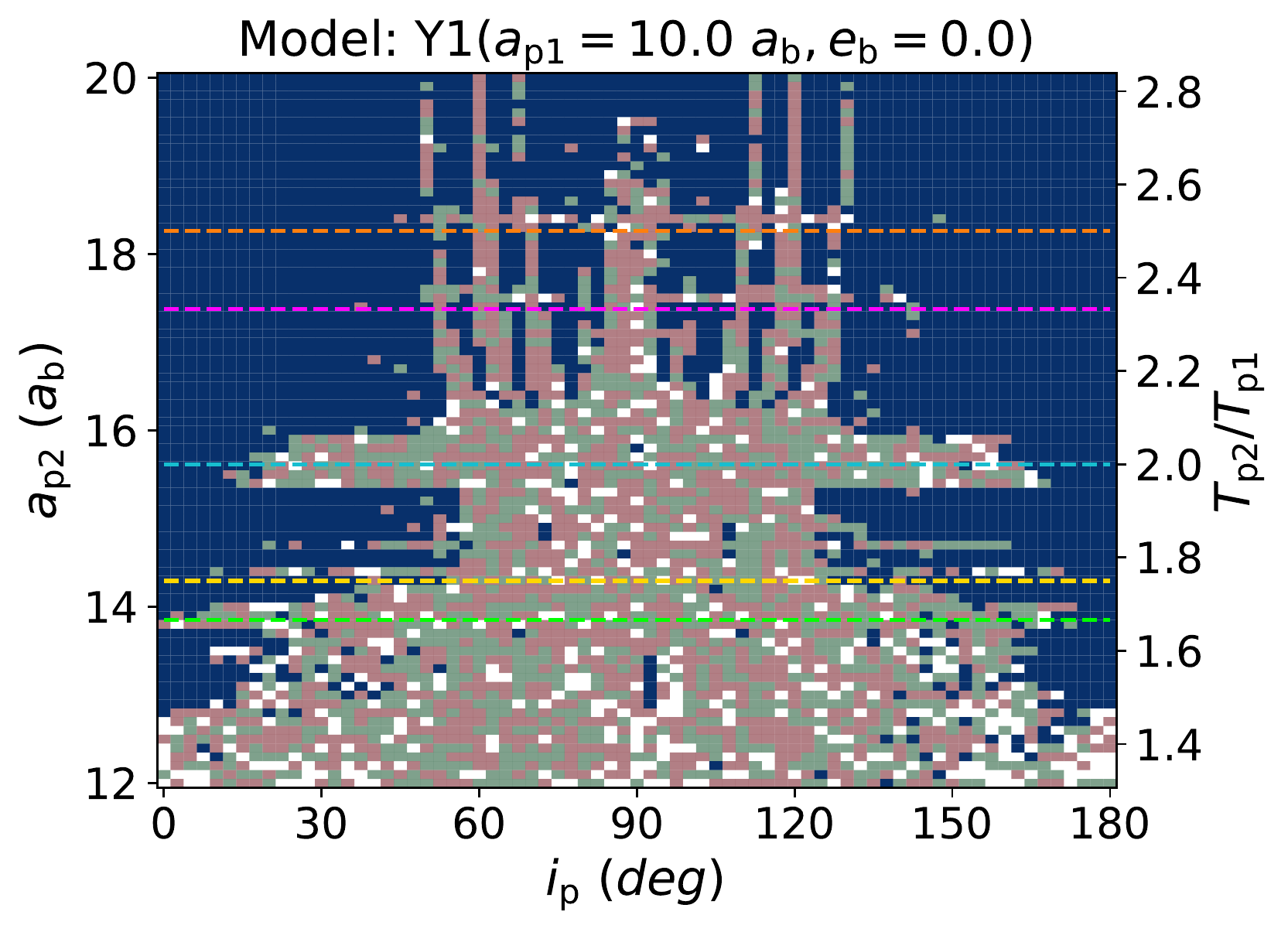}
    \includegraphics[width=8.7cm]{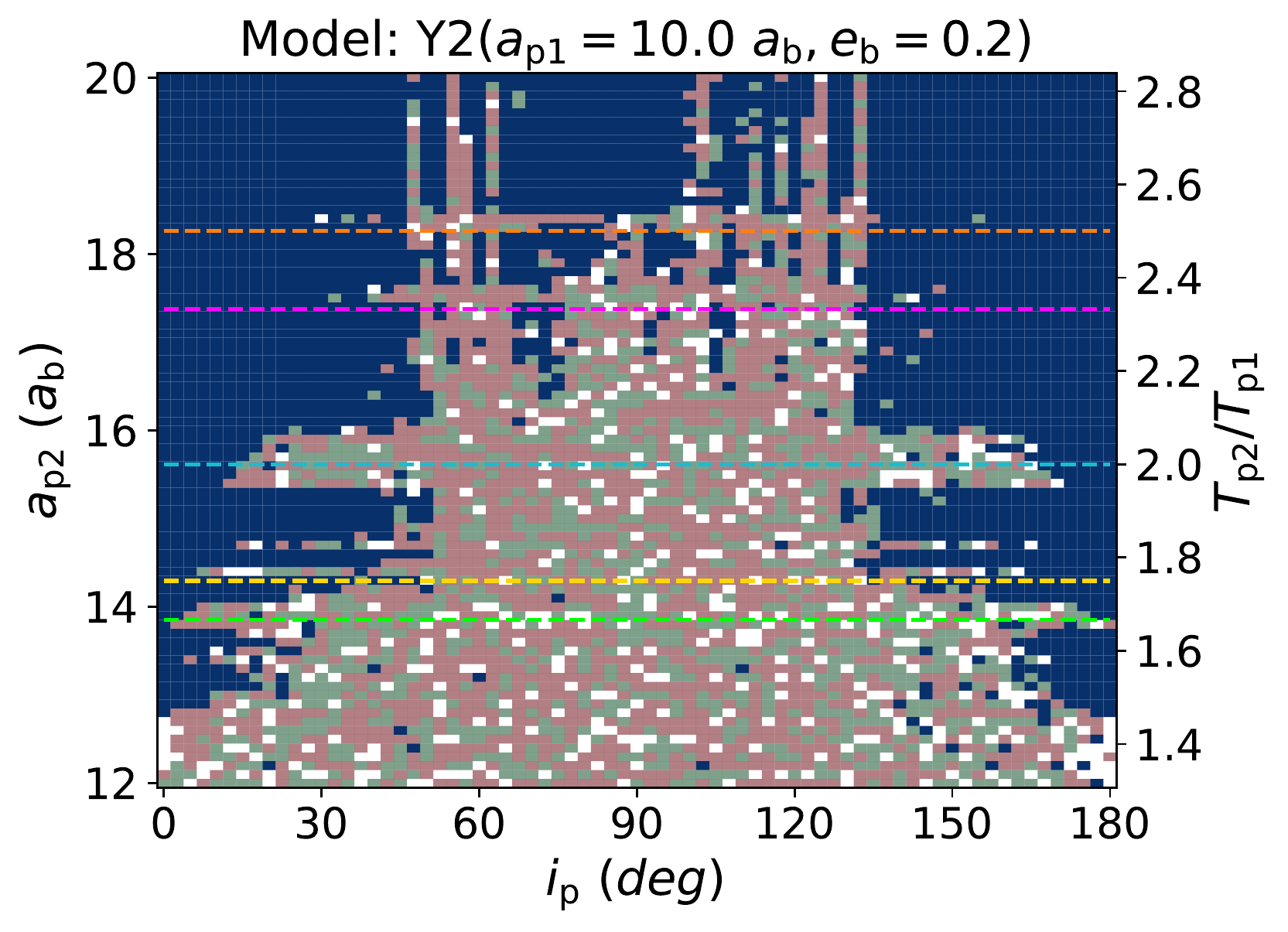}
    \includegraphics[width=8.7cm]{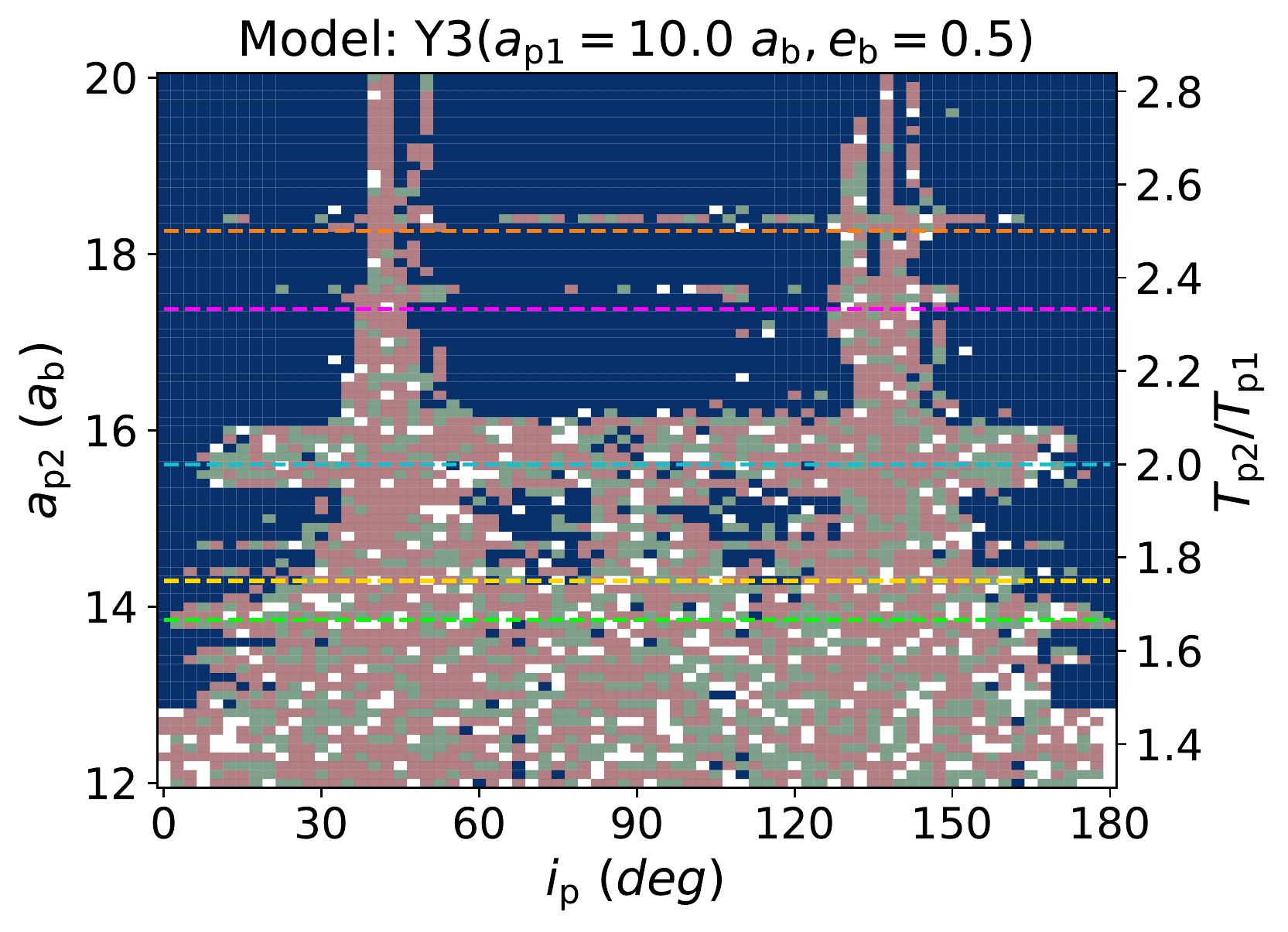}       
    \includegraphics[width=8.7cm]{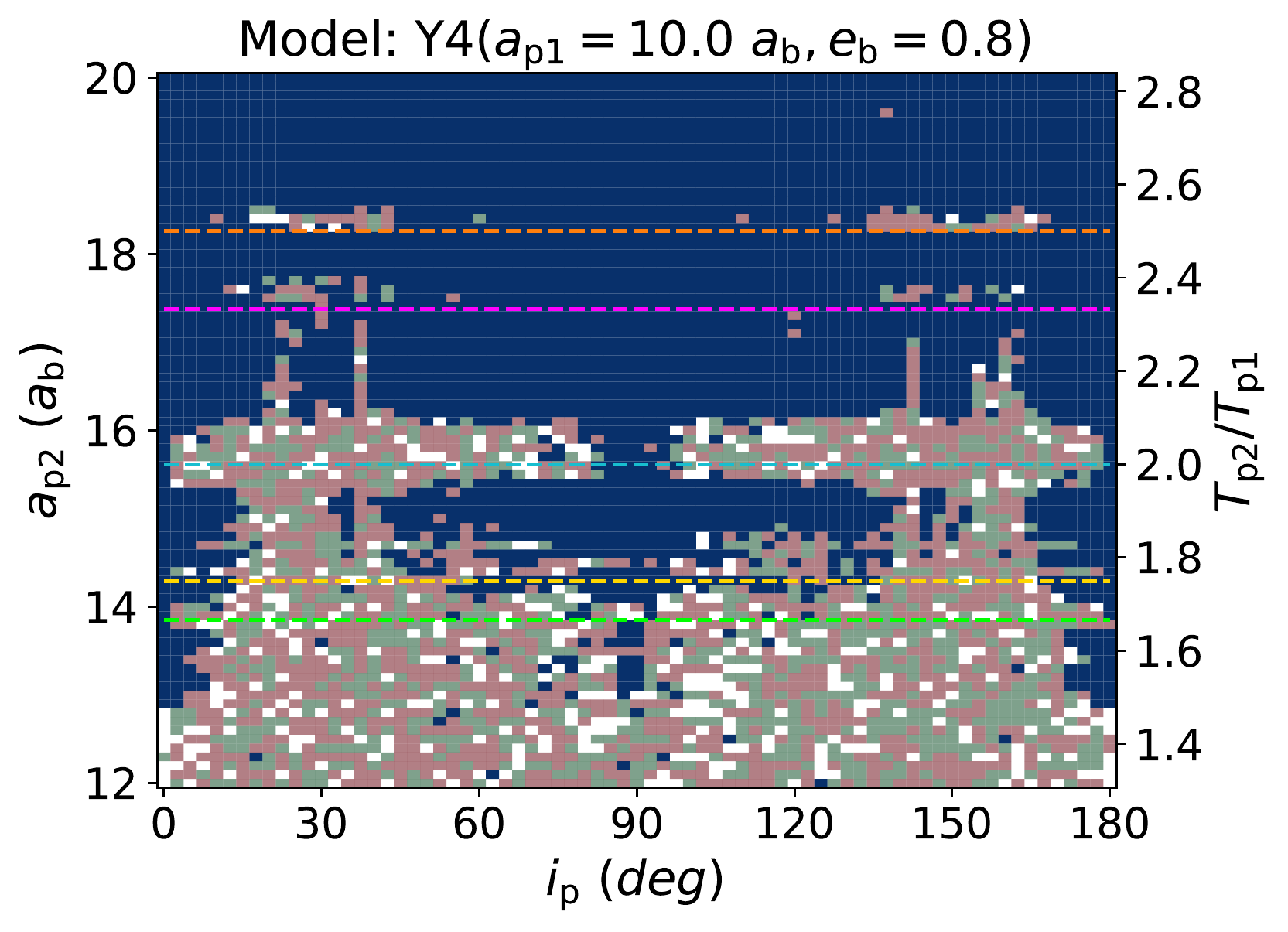}
    \caption{Same as Fig.~\ref{fig:mapa5}  except $a_{\rm p1}$ = 10$a_{\rm b}$.}
     \label{fig:mapa10}
\end{figure*}

\begin{figure*}
  \centering
    \includegraphics[width=8.7cm]{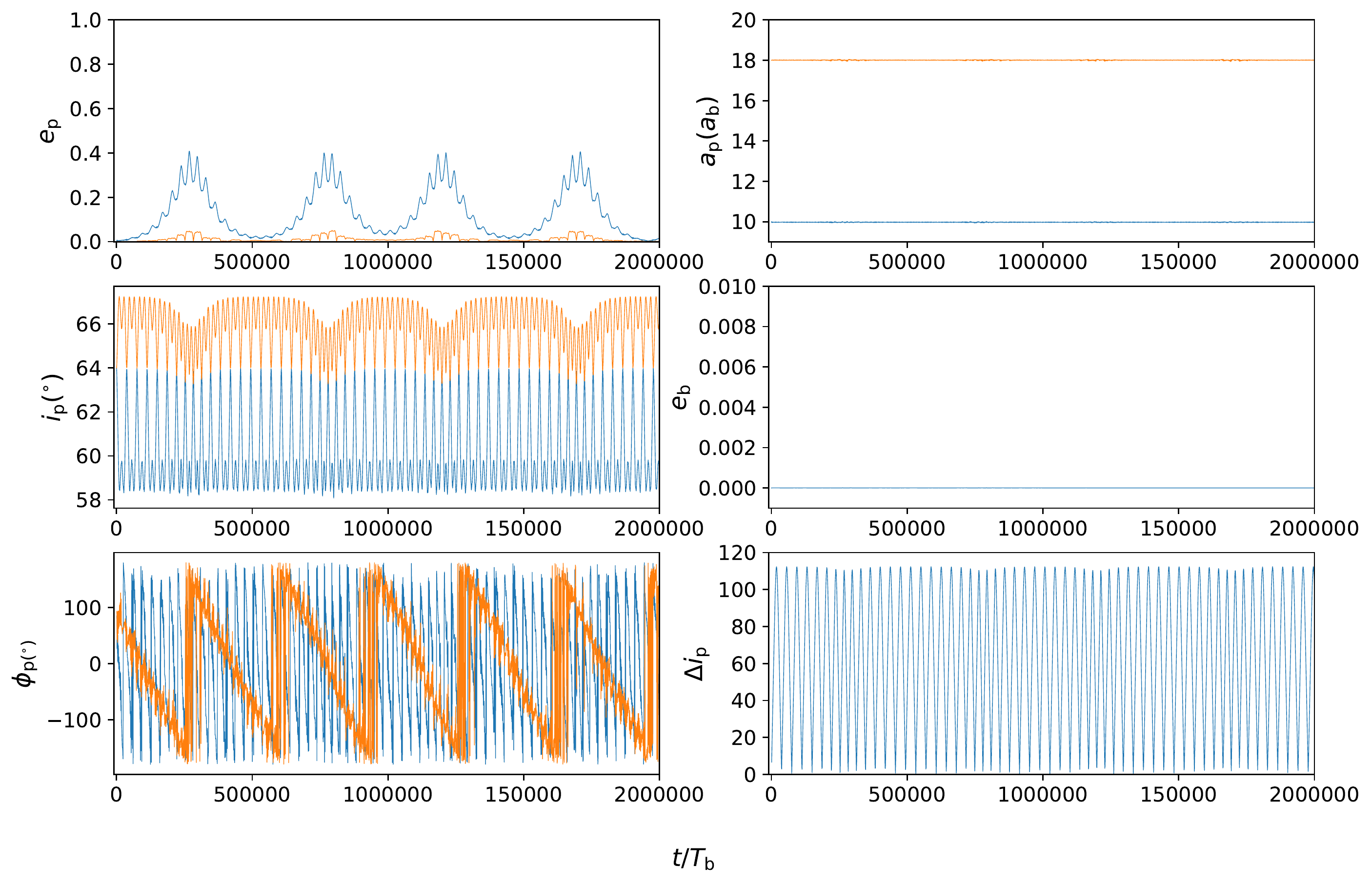}
    \includegraphics[width=8.7cm]{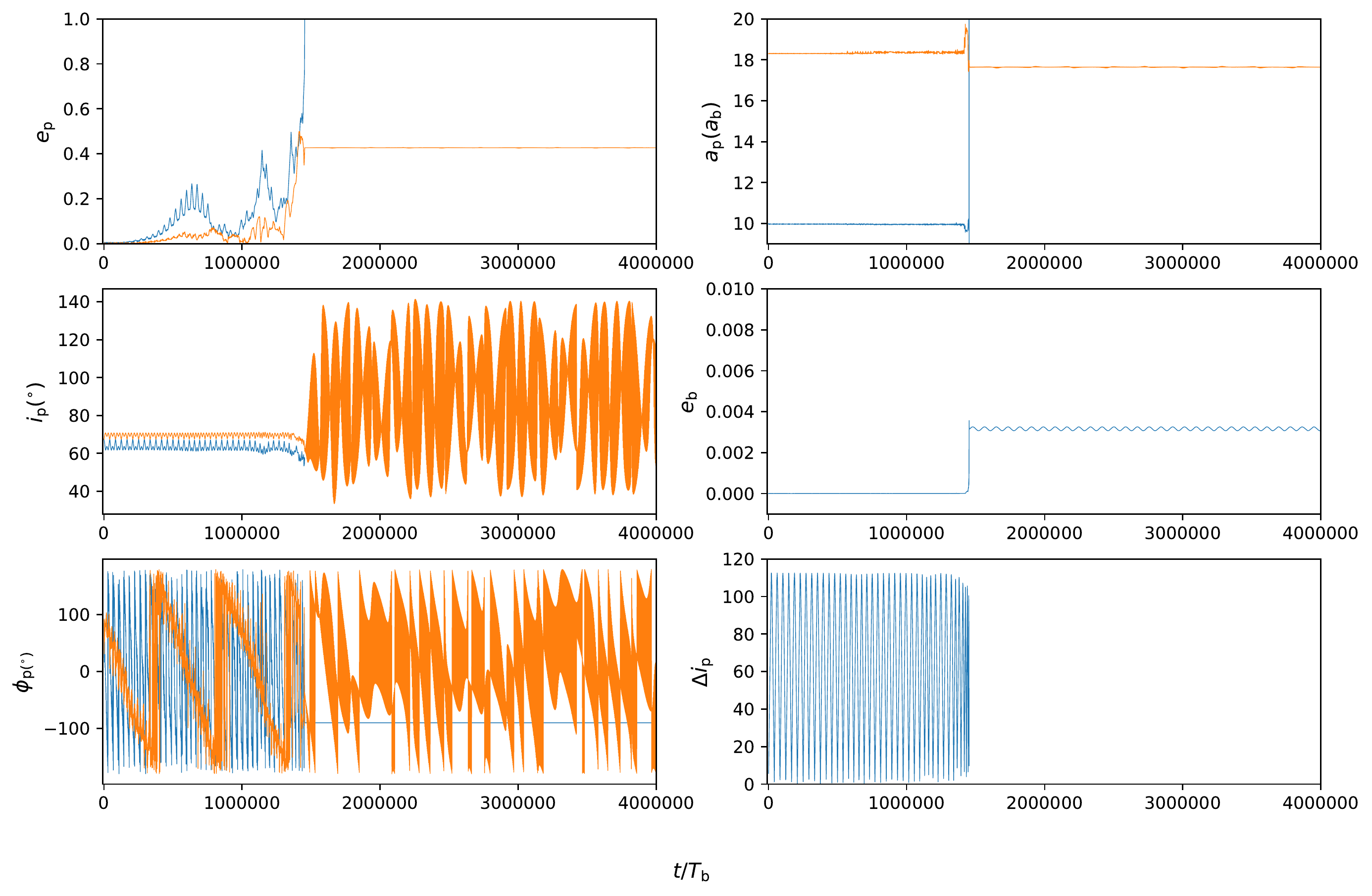}
    \caption{Examples of the particle dynamics of planets that undergo eccentricity oscillations in four-body simulations in model Y1 of Fig.~\ref{fig:mapa10} that has $e_{\rm b}=0$ and $a_{\rm p1}=10\,a_{\rm b}$.  Each set of six panels shows the planet eccentricity (top left), planet semi-major axis (top right), planet inclination (middle left), binary eccentricity (middle right), planet longitude of ascending node (lower left) and the mutual planet inclination (lower right). The initial planet properties are  $i_{\rm p} =64^{\circ}$ and $a_{\rm p2} = 18\, a_{\rm b}$ in the left panels and $i_{\rm p} = 67.5^{\circ}$ and $a_{\rm p2} = 18.3\,a_{\rm b}$ in right panels.  Each panel shows the  orbital evolution of the inner planet (blue line) and the outer planet (yellow line) except the two lower right panels that just show one line in each. The blue lines end in panels after the inner planet becomes unstable.}
     \label{fig:KL}
\end{figure*}

We now consider the case where the inner planet is farther out at $a_{\rm p1}  = 10\,a_{\rm b}$, models Y1--Y4.  Fig.~\ref{fig:mapa10} is the same as Fig.~\ref{fig:mapa5} except the initial $a_{\rm p1} = 10\,a_{\rm b}$ and the initial $a_{\rm p2}$ is in the range 12 -- 20$\,a_{\rm b}$. The integration time is increased to  $14 \times 10^6\ T_{\rm b}$ so that the inner planet has the same number of orbital periods as the simulations in the previous section with  $a_{\rm p1}=5\, a_{\rm b}$. We can then confirm whether the system is stable or not under the same physical timescale for each case.

The stability maps have changed dramatically, although there are the same resonance regions as those of in Fig.~\ref{fig:mapa5} including the 5:2, 7:3, 2:1, 7:4 and 5:3 resonances. Note  that the resonances in Fig.~\ref{fig:mapa10} are narrower than those  in Fig.~\ref{fig:mapa5} because the the range of $a_{\rm p2}$ is wider. The same feature can be seen in Fig.~\ref{fig:mapa20} next subsection which the range of $a_{\rm p2}$ is wider than Fig.~\ref{fig:mapa10}.

Between the 2:1 and 7:4 resonances, in contrast to the previous figure, the number of unstable orbits decreases as $e_{\rm b}$ increases. For $e_{\rm b}$ = 0.0 and 0.2, most of the CBPs can only be stable in this region if $i_{\rm p}$< 55$^\circ$ or > 125$^\circ$. Even though those planets are in prograde or retrograde circulating orbits, von Zeipel-Kozai-Lidov \citep[ZKL,][]{vonZeipel1910,Kozai1962,Lidov1962} oscillations may occur because the two planets are distant from the binary and each other. The classical ZKL mechanism occurs in a binary system that is perturbed by a distant third body. This dynamical effect causes the small object in the binary system to undergo oscillation of the argument of  periapsis, which results in a periodic exchange between its eccentricity and inclination. The  critical minimum tilt angle is 39.2$^\circ$.  However, our system is not in the hierarchical triple with a test particle limit.  In this case, $e_{\rm p1}$ and $e_{\rm p2}$ can get excited, and thus the CBPs can undergo a close encounter. 

The eccentricity of a single circumbinary planet remains constant in time to quadrupole order
of the binary potential \citep[e.g.,][]{Farago2010}. If the binary is replaced by a single star, then the two planets that start mutually coplanar and on circular orbits would remain coplanar and on circular orbits for all but the closest separations in the stability maps. So neither case results in eccentricity growth. Instead, we suggest that a modified form of ZKL may be acting as the mutual
inclination of the planets increases and oscillates due to their different nodal precession rates caused by the central binary, as discuss below.

In Fig~\ref{fig:KL}, we show the dynamics of one survivor case (left) and one unstable case (right) for systems that undergo eccentricity oscillations. For each set of six panels, the upper-left panel is $e_{\rm p}$, the upper-right is $a_{\rm p}$, the middle-left panel is $i_{\rm p}$, the middle-right panel is $e_{\rm b}$, the lower-left panel is $\phi_{\rm p}$ and the lower-right panel is the mutual inclination between two planets, $\Delta i_{\rm p}$, defined by Equation (\ref{Deltaip}). The blue lines represent the inner planet while the yellow lines represent the outer planet except in the two lower right panels where there is only one line. Both cases show eccentricity oscillations of the inner planet similar to ZKL oscillations.

However, the mutual inclination between the planets $\Delta i_{\rm p}$ undergoes rapid large oscillations of nearly constant amplitude  on a timescale that is much shorter than for the eccentricity oscillations, unlike the standard ZKL oscillations. These rapid oscillations are the result of the planets' relative precession at nearly constant tilt $i_{\rm p}$ relative to the binary. This behaviour is quite different from the standard ZKL mechanism in which the inclination and eccentricity oscillation periods are the same  and the eccentricity varies a function of the inclination. The time-varying misalignment might result in ZKL-like oscillations. We note also that there is some evidence of long period inclination oscillations of the outer planet relative to the binary in the left middle panel of Fig~\ref{fig:KL}. The mechanism
should be investigated further.

For the case of small $e_{\rm b}$ = 0.2 (model Y2), considering planets closer in than the 2:1 resonance, there is still a stable range of $i_{\rm p} < 55^\circ$ or > 125$^\circ$ and panel is quite similar to model Y1. However, for $e_{\rm b}$ = 0.5 (model Y3), the unstable region increases to between 30$^\circ < i_{\rm p}< 150^\circ$. This might be because of nodal oscillations driven by the binary since the boundary of unstable and stable orbits is close to the critical inclination for nodal libration. Unlike the previous two models, there are some stable orbits around the polar librating region which implies that the two planets may be stable if they undergo strong nodal libration with the binary and the nodal oscillation between the binary and planet could offset the effect of the ZKL oscillation. 

This effect is more significant for $e_{\rm b}$ = 0.8 (model Y4). The two planets closer in  than the 2:1 resonance can be not only stable in the coplanar orbits but also in the polar librating orbits.  This is because the secular resonance with the binary dominates the system and the ZKL mechanism cannot operate. However, when $i_{\rm p}$ is away from the stationary inclination, it can vary a lot in the polar librating orbits \citep[see][for details]{Chen20192}. This behaviour may destabilise the orbits of the two planets. 

Beyond the 2:1 resonance, the maps with $e_{\rm b}$ = 0.0 and 0.2 have complex orbital stability. Moreover, a massive planet can exchange angular momentum between its inclination and the binary eccentricity \citep{Chen20192}. A two CBP system can have significant angular momentum to exchange with the binary. The different libration timescales of the two planets can drive a complicated $e_{\rm b}$ oscillation. We find that $e_{\rm b}$ in model Y1, which initially is 0, grows to significant values (it can reach up to 0.1 in some pixels). Thus, a CBP around $i_{\rm p}=90^\circ$ in this system may undergo a polar libration with the binary. Considering the diverse orbital interactions we mentioned above and combining with resonances, the complicated orbital stability regions beyond the 2:1 resonance can be seen in both $e_{\rm b}$ = 0.0 and 0.2. 

On the contrary, the maps with $e_{\rm b}$ = 0.5 and 0.8 are much more stable than the other two maps beyond the 2:1 resonance except in small resonance regions. The stable orbits in this region can be explained by the nodal oscillations between the binary and the outer planet which has more angular momentum to exchange dominating interactions in these systems. It dominates the ZKL mechanism between the two planets. 
Models H1 and H2 in  Fig.~13 in \citet{Chen2022} display two CBPs with $i_{\rm p} = 80^\circ$ and 90$^\circ$ at 10$\,a_{\rm b}$ and at 18$\,a_{\rm b}$ in a binary system with $e_{\rm b} = 0.5$. In these two panels, the two planets undergo nodal libration with the binary and the tilt oscillations between the two planets are tiny. This result is similar to what we see in the stability maps of model Y3 and Y4. Additionally, the map with $e_{\rm b} = 0.5$ has the same unstable regions but they are much wider than those in Fig.~\ref{fig:mapa5}.

In contrast to the $a_{\rm p1}= 5\,a_{\rm b}$ maps, the ratio $R_{\rm i/o}$ does not vary a lot with increasing $e_{\rm b}$ because the two planets are farther away from the binary, fewer planets can reach the stability limit of the binary. These ratios are close to $R_{\rm i/o}=4/5$ (model Y1), $2/3$ (model Y2), $1/2$ (model Y3) and $7/10$ (model Y4). 

\subsection{Binary system with two planets: $a_{\rm p1}$ = 20$\,a_{\rm b}$}

\begin{figure*}
  \centering
    \includegraphics[width=8.7cm]{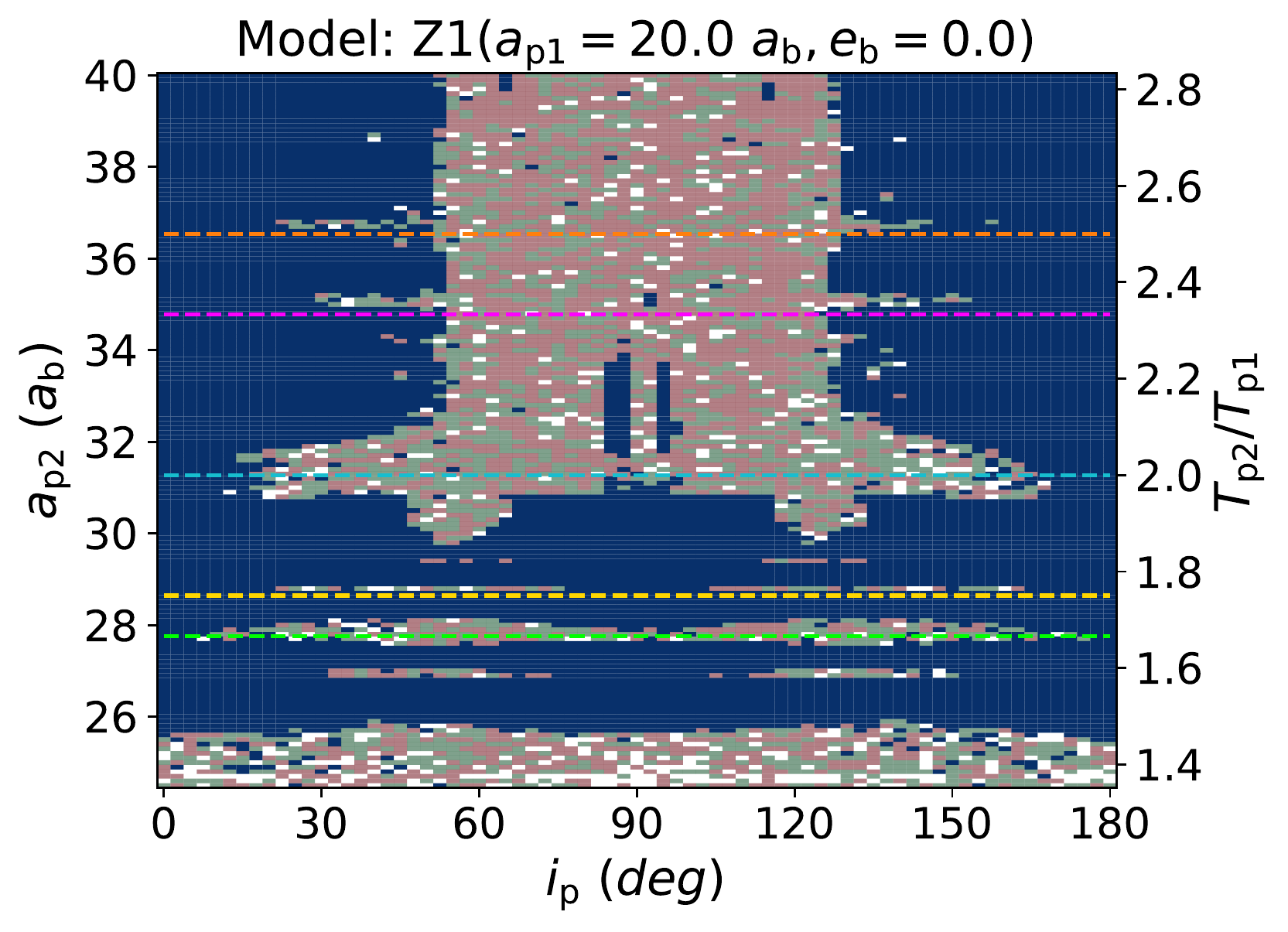}
    \includegraphics[width=8.7cm]{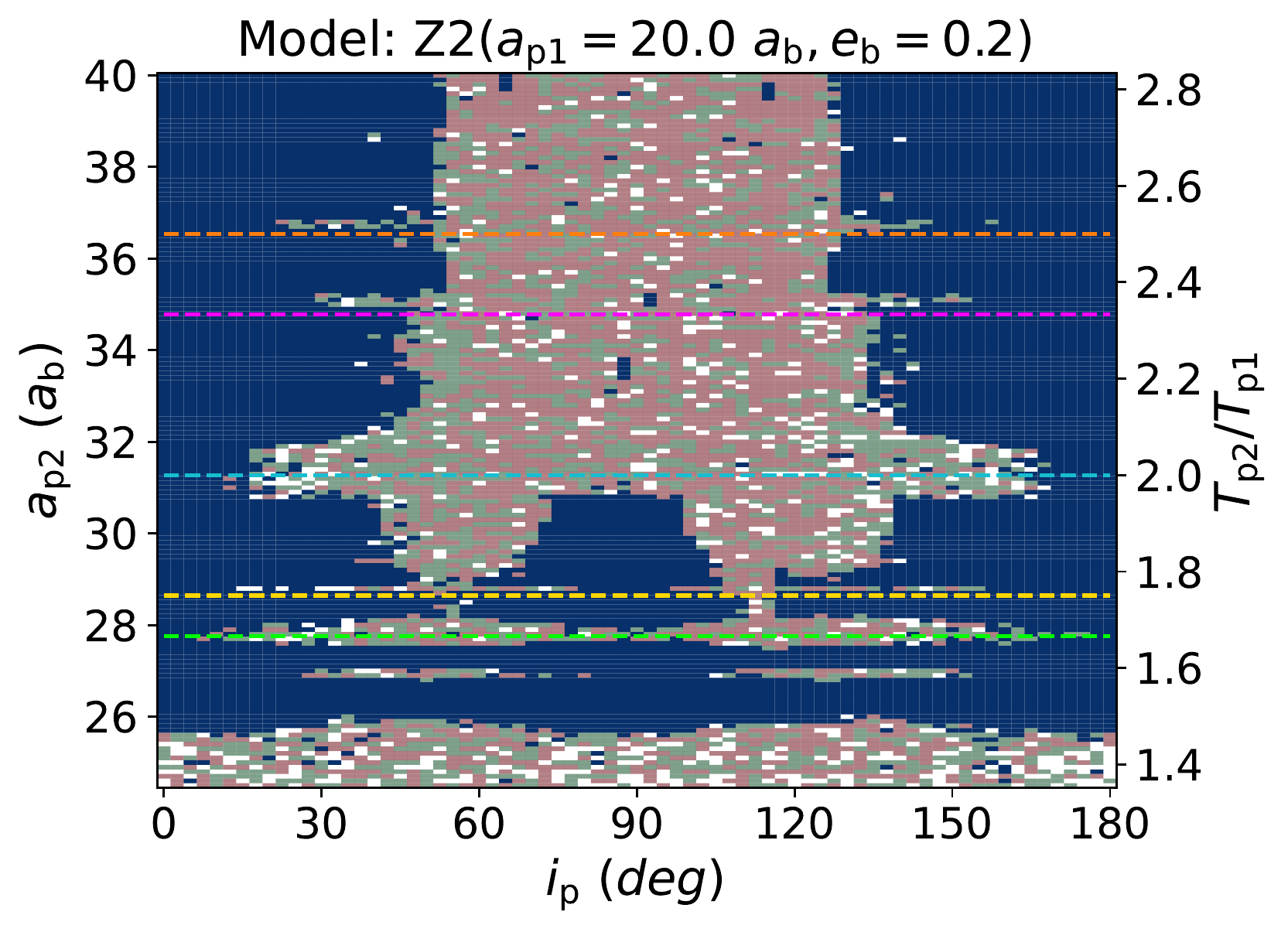}
    \includegraphics[width=8.7cm]{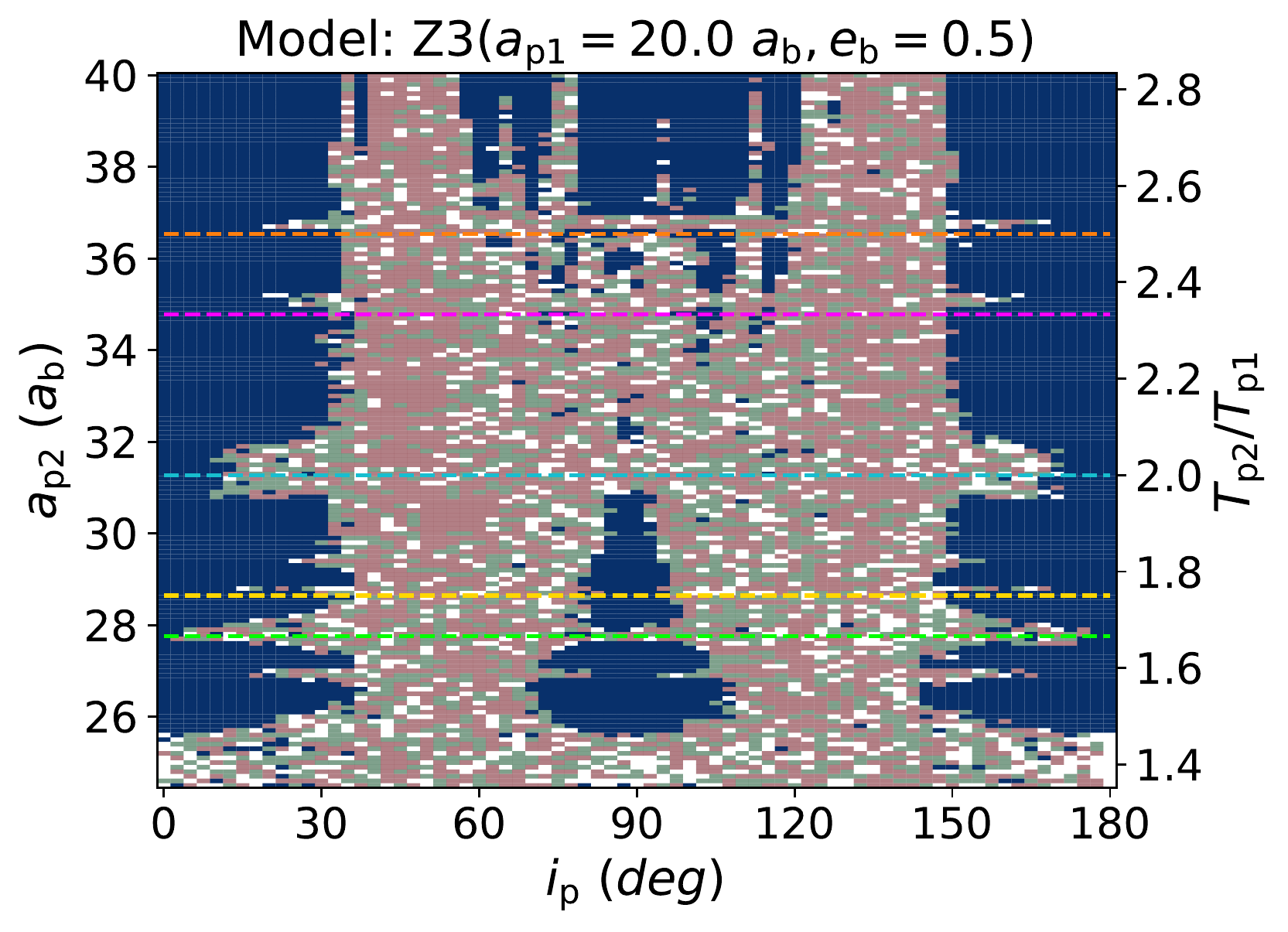}       
    \includegraphics[width=8.7cm]{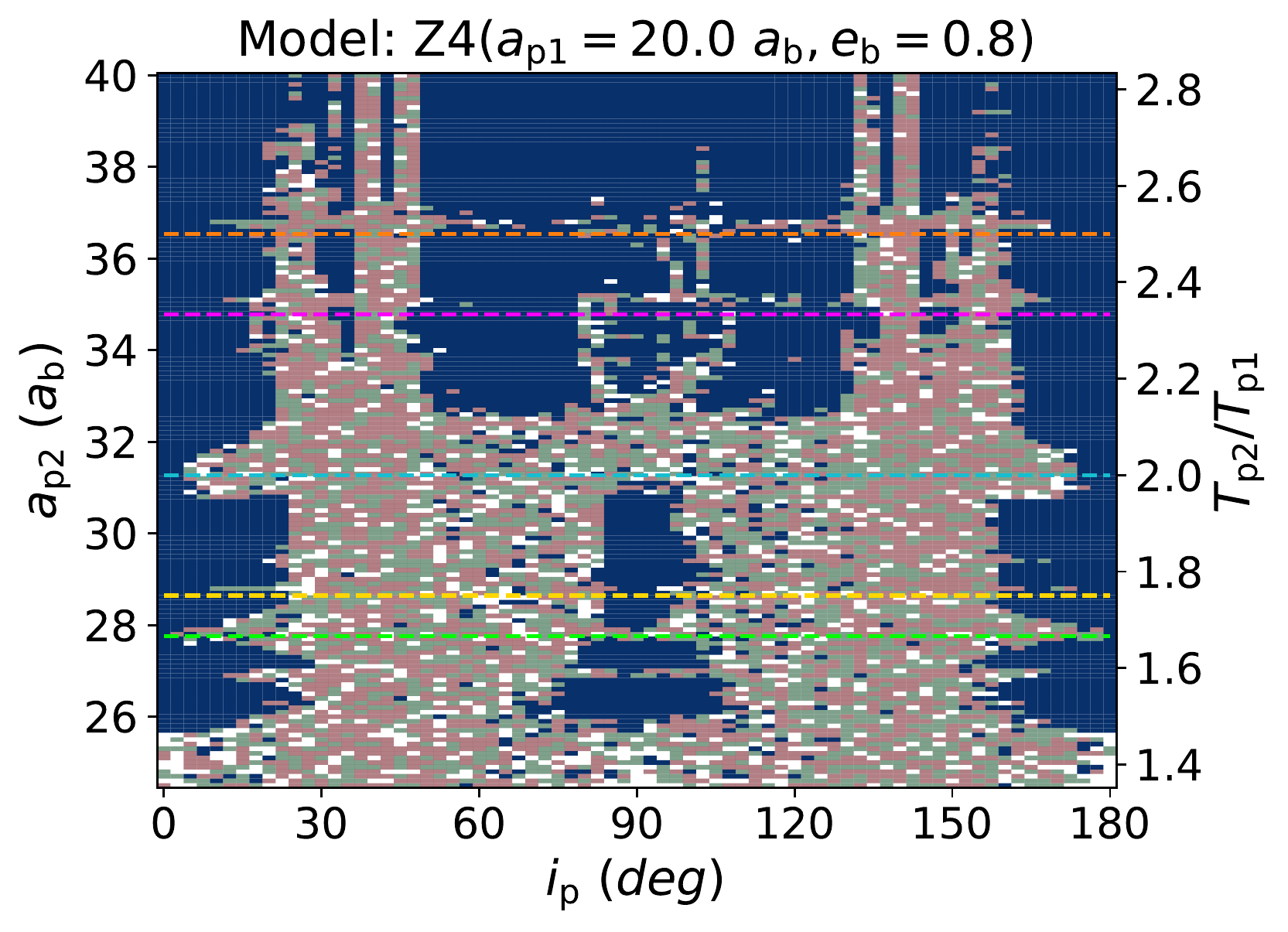}
    \caption{Same as Fig.~\ref{fig:mapa5} except the inner planet is at $a_{\rm p1}$ = 20$a_{\rm b}$.}
     \label{fig:mapa20}
\end{figure*}

\begin{figure}
    \centering
    \includegraphics[width=8.7cm]{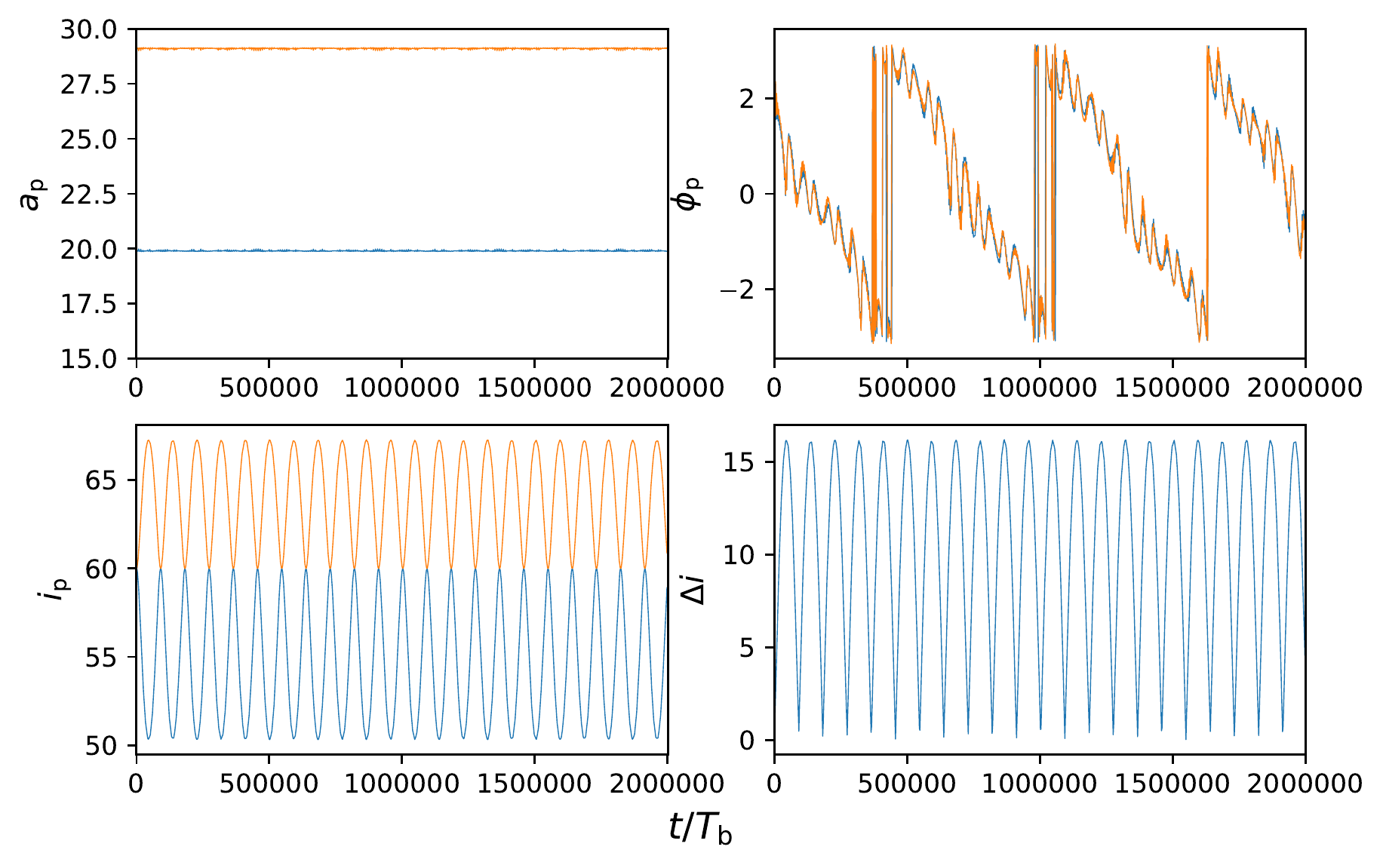}
    \caption{The orbital evolution of two CBPs in model Z1 that has $e_{\rm b}=0$.  The two planets are initially located at $a_{\rm p1}$ = 20 $a_{\rm b}$, $a_{\rm p2}$ = 29 and have $i_{\rm p}$ = 60$^\circ$. The upper left panel shows their semi-major axes, the upper right panel shows their longitude of ascending nodes and the lower left panel shows their inclinations. In each of these, the blue line shows the inner planet and the yellow line shows the outer planet. The lower right panel shows the mutual inclination of the planets. }
    \label{fig:lock}
\end{figure}

\begin{figure*}
  \centering
    \includegraphics[width=8.7cm]{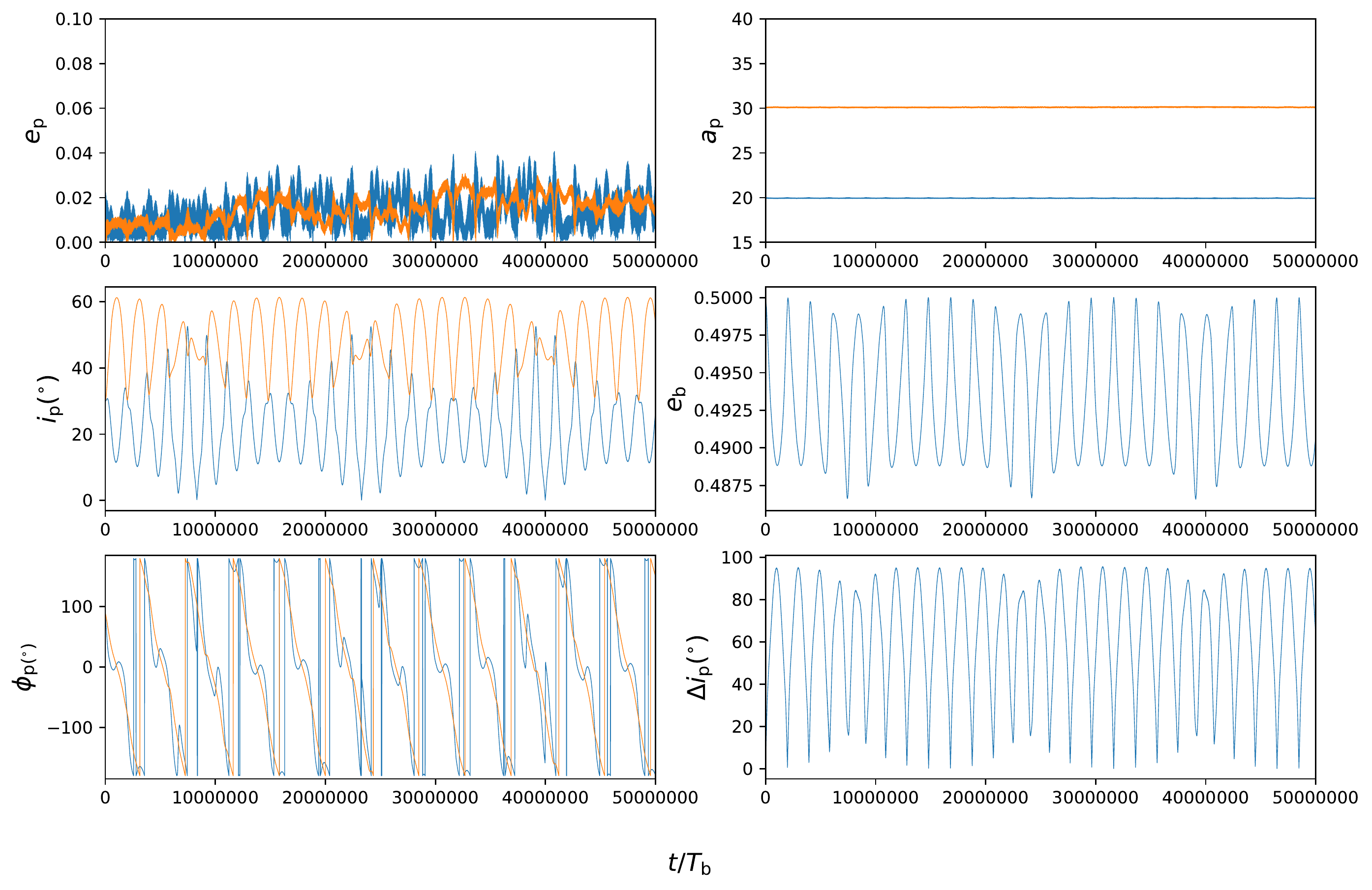}
    \includegraphics[width=8.7cm]{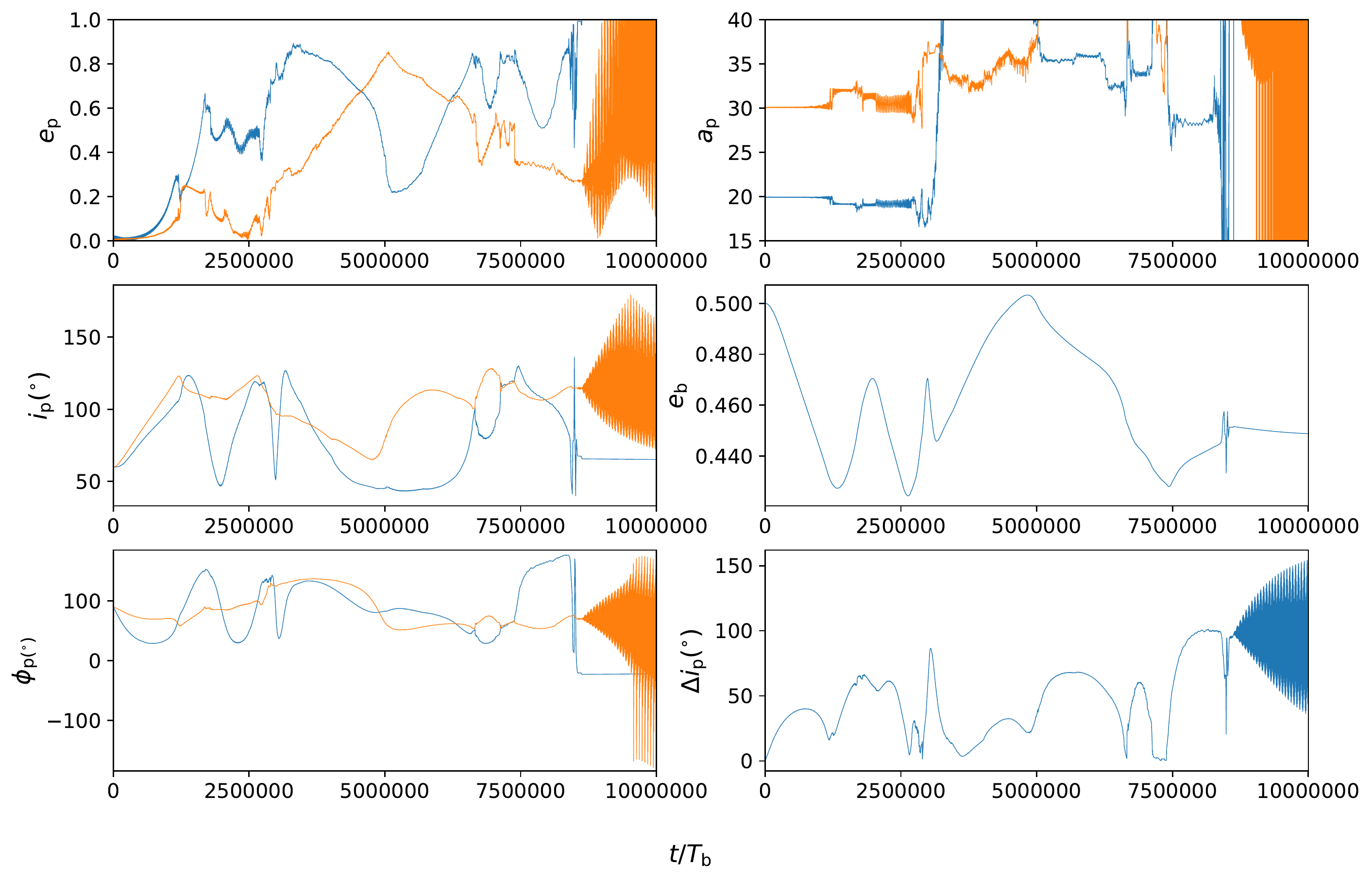}
    \includegraphics[width=8.7cm]{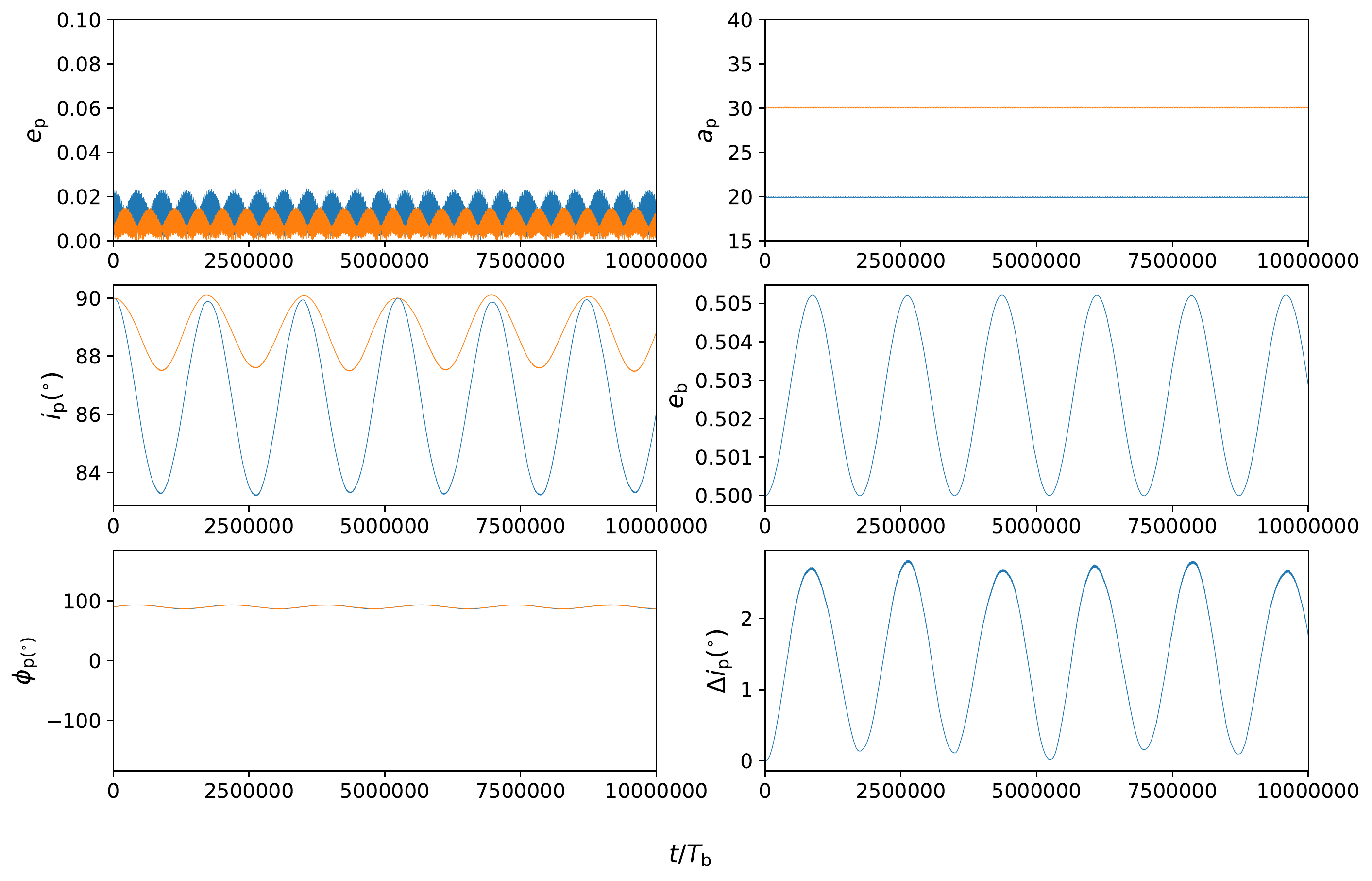}       
    \caption{Orbital evolution of two CBP cases with an initial $a_{\rm p1} = 20\,a_{\rm b}$, $a_{\rm p2} = 30\,a_{\rm b}$ and initial $i_{\rm p}=30^{\circ}$ (upper-left), $60^{\circ}$ (upper-right) and $90^{\circ}$ (lower) in model Z3 with $e_{\rm b}=0.5$. Each set of six panels are the same as those described in Fig.~\ref{fig:KL}.}
     \label{fig:map306090}
\end{figure*}

Fig.~\ref{fig:mapa20} is similar to Figs.~\ref{fig:mapa5} and~\ref{fig:mapa10} but with initial $a_{\rm p1} = 20\,a_{\rm b}$ and  initial $a_{\rm p2}$ in the range 25 -- 40$\,a_{\rm b}$. The integration time is  $40 \times 10^6\ T_{\rm b}$. The stability maps of models Z1 -- Z4 are quite different from the previous simulations with the smaller orbital radii of the inner planet. We see the same resonance regions including the 5:2, 7:3, 2:1, 7:4 and 5:3 resonances and there are additionally two small resonances which are below the 5:3 resonance and above the 7:4 resonance. 

For the maps with $e_{\rm b}$ = 0.0 and 0.2 (models Z1 and Z2), the two upper panels are unlike the previous maps with the same $e_{\rm b}$. The orbits interior to the 2:1 resonance are quite stable for $a_{\rm p2} \lesssim 25.6 \,a_{\rm b}$, apart from narrow resonance regions. Since the two planets are close to each other and farther away from the binary than the $a_{\rm p1}=5 a_{\rm b}$ and $10 a_{\rm b}$ cases, the two planets undergo mutual libration that limits their relative inclination. Thus their mutual inclination is small enough to prevent ZKL oscillations and keep them stable. In Fig~\ref{fig:lock}, we show the dynamics of one simulation from model Z1 with initial $a_{\rm p1}$ = 20 $a_{\rm b}$, $a_{\rm p2}$ = 29 $a_{\rm b}$ and $i_{\rm p}$ = 60$^\circ$. In the upper-right panel, we see that their phase angles are almost locked and thus their largest $\Delta i$ (lower-right panel) is about 15$^\circ$ during the tilt oscillations. On the other hand, between $i_{\rm p} = 55^\circ$ and 125$^\circ$, the ZKL mechanism dominates the system beyond the 2:1 resonance because the planets undergo mutual circulation with respect to the binary and their $\Delta i$ is larger than the critical ZKL angle. Consequently, $e_{\rm p}$ of the two planets get excited and they go unstable due to  close encounters with each other.

For the maps with $e_{\rm b}$ = 0.5 and 0.8 (model Z3 and Z4), the two lower panels have a more similar configuration to the previous plots that have initial  $a_{\rm p1}$ = 5$\,a_{\rm b}$ and $a_{\rm p1}$ = 10$\, a_{\rm b}$. The orbits of the outer planets which are interior to the 2:1 resonance are only stable for $i_{\rm p}$ < 30$^\circ$ or > 150$^\circ$ and around the polar orbit region. The large unstable region is caused by large variations of $i_{\rm p}$ due to the nodal oscillation with the binary. In Fig.~\ref{fig:map306090}, we show the orbital evolution of two planets from model Z3 which are initially located at $a_{\rm p1} = 20\,a_{\rm b}$ and $a_{\rm p2} = 30\,a_{\rm b}$ and their initial $i_{\rm p} = 30^\circ$ (upper-left), 60$^\circ$ (upper-right) and 90$^\circ$ (lower). For the $i_{\rm p}=30^\circ$ case, two planets not only undergo the tilt oscillation with each other but also undergo the nodal oscillation with the binary individually. This unusual behaviour has been found in our previous study (see Figure 6 in \citealt{Chen2022}.) For the $i_{\rm p}=60^\circ$ case, $e_{\rm p}$ get excited in a short time and it results in complicated interactions. For the $i_{\rm p}=30^\circ$ and 90$^\circ$  cases, the two planets under nodal circulation (libration) with the binary and no ZKL mechanism is triggered in these two cases. The variation of $i_{\rm p}$ is small if the planets inclination is close to the stationary inclination. Hence, the orbits close to coplanar and polar configurations tend to be stable. 

Beyond the the 2:1 resonance, more planets can be stable with increasing $e_{\rm b}$, so model Z3 has less stable orbits than model Z4 in this region. The strong nodal oscillations between the planet and the binary dominates the system no matter whether $\phi_{\rm p}$ are locked or not.  Thus, the effect of the ZKL mechanism is not obvious in this region. In contrast to the $a_{\rm p1} = 5\,a_{\rm b}$ maps but similar to the $a_{\rm p1} = 10.0\,a_{\rm b}$ maps, $R_{\rm i/o}$ does not vary a lot with increasing $e_{\rm b}$. These ratios are close to $R_{\rm i/o}=1$ (model Z1), $4/5$ (model Z2), $7/10$ (model Z3) and $1$ (model Z4).

\section{Final distributions of  the stable planets}
\label{sing}

We now consider the orbital properties of the planets that remain stable at the end of our simulations. We consider histograms of the final planet eccentricity and the final planet semi-major axis relative to its initial semi-major axis. Each histogram represents the sum of both the inner and outer planets since the differences between the distributions for the inner and outer planets are small, especially when the planets are far from the binary.

\begin{figure*}
  \centering

    \includegraphics[width=17.4cm]{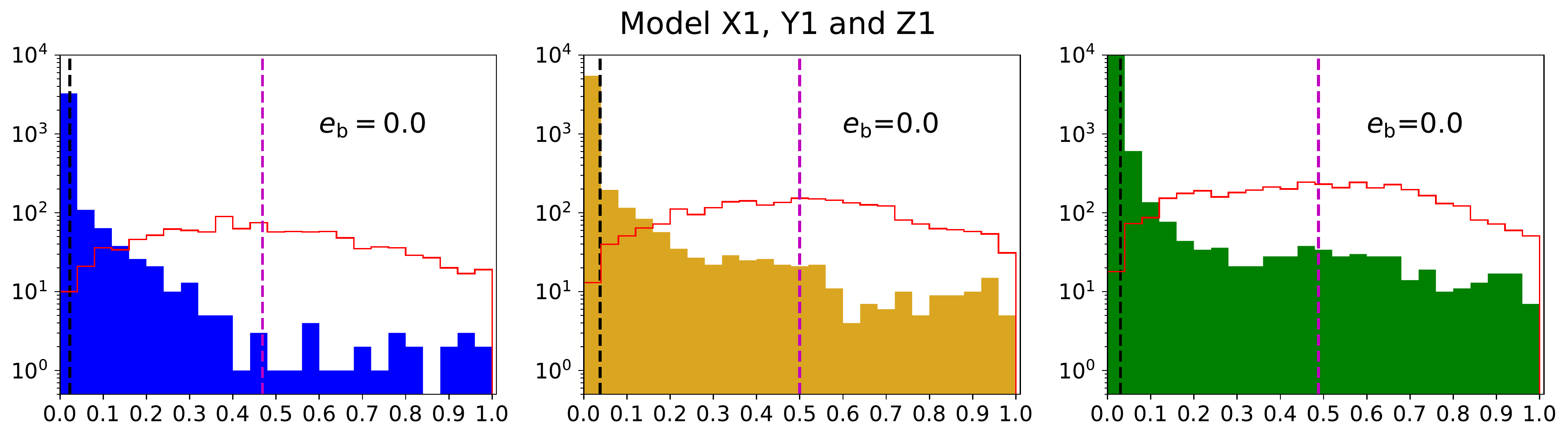}
    \includegraphics[width=17.4cm]{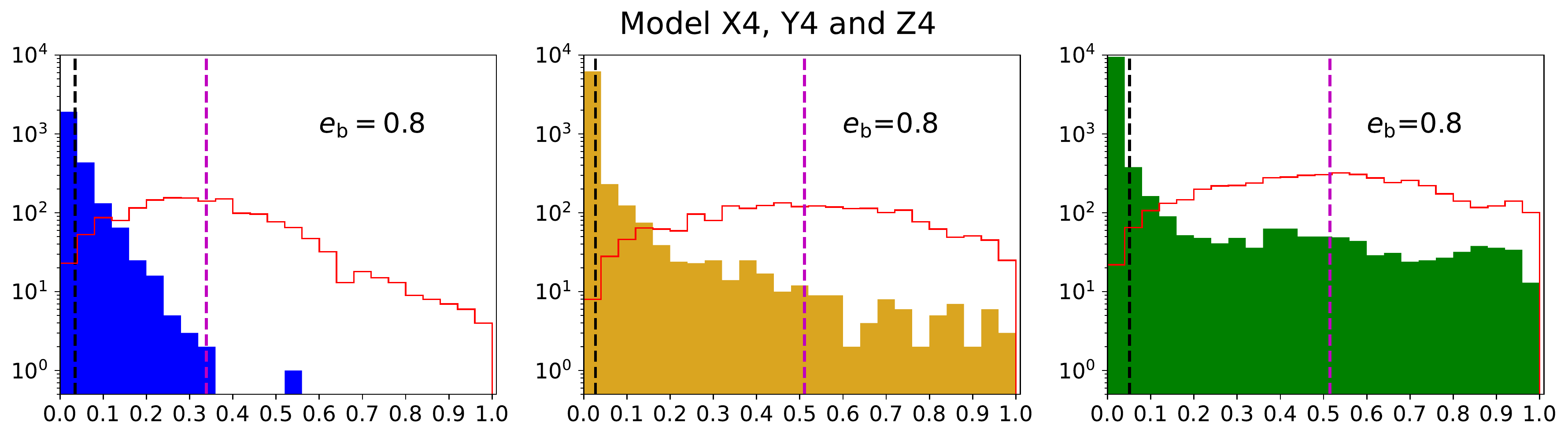}
    
    \textbf{$e_{\rm p}$}

    \caption{Histograms of the final $e_{\rm p}$ distributions. The filled histograms show cases where both planets survive and the hollow histograms show  cases where only one planet survives in models X1, Y1 and Z1 (upper panels) and models X4, Y4 and Z4 (lower panels). The $x$-axis is $e_{\rm p}$ and the $y$-axis is the log of the number of planets. The black dashed line is the mean value of $e_{\rm p}$ of the filled histogram while the purple dashed line is mean value of $e_{\rm p}$ of the red hollow histogram.}
     \label{fig:ee}
\end{figure*}

\begin{figure*}
  \centering
    \includegraphics[width=17.4cm]{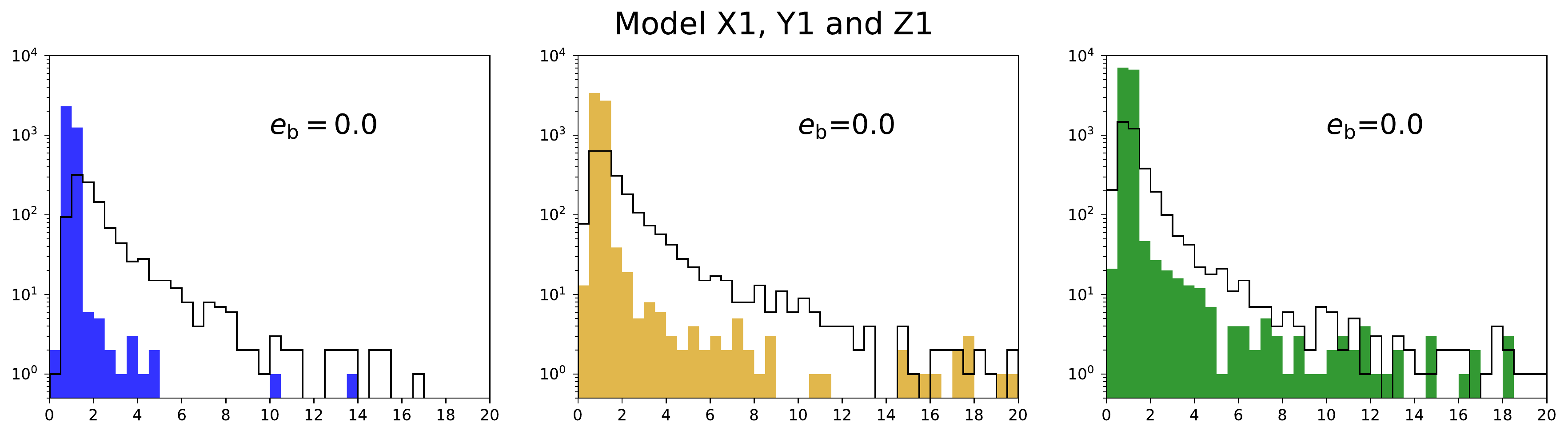}
    \includegraphics[width=17.4cm]{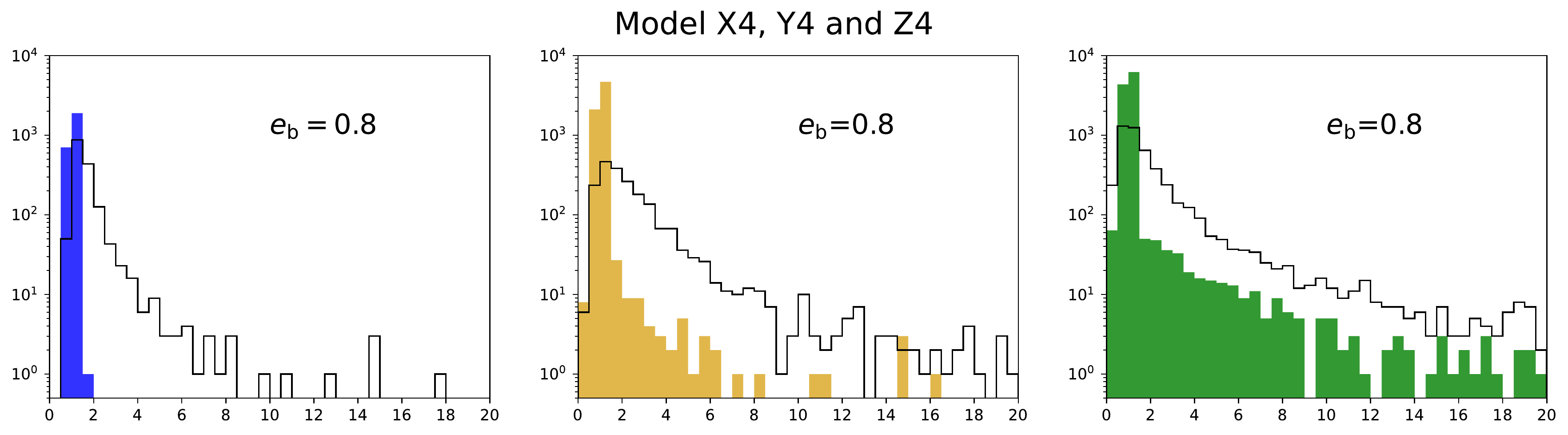}
    \text{Final $a_{\rm p}$/ initial $a_{\rm p}$} 
    \caption{Histograms of the final $a_{\rm p}$/initial $a_{\rm p}$  distributions. The filled histograms show cases where both planets survived and the black hollow histograms show cases where only one planet survived in models X1, Y1 and Z1 (upper panels) and models X4, Y4 and Z4 (lower panels). The $x$-axis is the final $a_{\rm p}$/ the initial $a_{\rm p}$ and $y$-axis is the log of the number of planets.}
     \label{fig:aa}
\end{figure*}

\subsection{Two stable planets cases}

For the range of parameters that we have considered, the majority of the outcomes in our stability maps are two stable planets. To understand the orbital dynamics of the two planets, in Fig.~\ref{fig:ee}, we plot histograms of the final $e_{\rm p}$ of the models with $e_{\rm b} = 0.0$ (upper panels) and 0.8 (lower panels) in the filled coloured columns. The black dashed lines indicate the mean value of each model. The distribution and the mean values are not very sensitive to $e_{\rm b}$ and so we only display the models with $e_{\rm b} = 0.0$ and 0.8. The cases with a high final $e_{\rm p}$  are the result of planet-planet resonances \citep{Chiang2002a} and eccentricity oscillations (as shown in Fig~\ref{fig:KL}). Because the planets in models X1 and X4 are close to the binary, the maximum $e_{\rm p}$ attained is restricted by $e_{\rm b}$ because an eccentric CBP can be disturbed frequently by an eccentric binary. Thus, the final $e_{\rm p}$ in model X4 cannot exceed about 0.3. In general, the mean eccentricity of the planets is low in systems in which both planets remain stable.

Similar features can be seen in the histograms of the final $a_{\rm p}$ (here after $a_{\rm pf}$) scaled to the initial $a_{\rm p}$ (here after $a_{\rm p0}$)   in Fig~\ref{fig:aa}.  The filled coloured columns are the histograms of $a_{\rm pf}/a_{\rm p0}$ for models with $e_{\rm b} = 0.0$ (upper panels) and 0.8 (lower panels) for systems in which both planets remain stable. The values of $a_{\rm pf}/a_{\rm p0}$ are concentrated around 1 for all models. With a larger initial $a_{\rm p1}$, there are more planets that have  larger $a_{\rm pf}/a_{\rm p0}$ values, except in models X1 and X4. There are no planets that have $a_{\rm pf}/a_{\rm p0} > 2$ in model X4 because close encounters with the eccentric  binary results in destabilising the orbit of a planet.

\subsection{One planet survivor cases}

When a planetary system is unstable, the most likely outcome is that one planet remains as a survivor.
The probability of the outer planet remaining stable (red pixels) generally increases with increasing $e_{\rm b}$ (the lower-left panel in Fig.~\ref{fig:mapa20} is an exception).  The orbital parameters of the remaining CBP may change significantly  during the ejection.  The red hollow histograms in Fig.~\ref{fig:ee} represent the final distributions of $e_{\rm p}$ in cases where one planet survives. The purple dashed lines indicate the mean values of $e_{\rm p}$. The mean planet eccentricity is increased to about 0.5, except for model X4 which has a maximum of about $ 0.3$ because of close encounters with the binary. The large mean values of $e_{\rm p}$ in the surviving planets implies a strong interaction between the two planets occurs before one planet got ejected from the system.

The distributions of $a_{\rm pf}/a_{\rm p0}$ for the one survivor planet cases are shown in the black lines in Fig~\ref{fig:aa}. The trends are similar to the two stable planets cases.  Although most of the surviving planets have a final semi-major axis in the range $1\sim 2\, a_{\rm p0}$, a larger fraction of surviving planets have larger $a_{\rm pf}$ at the end of simulations compared to the two planet stable cases. This is a result of the strong interactions with the ejected planets.

\section{Discussion and Conclusions} 
\label{diss}

In this paper, we have studied the orbital stability of two misaligned CBPs with masses of 0.001 times the binary mass around a circular or an eccentric binary. In addition to the inclinations of planets, we also considered the influence of the semi-major axis of the inner planet, the separation of the two planets and the binary eccentricity. Unlike a normal multi-planetary system around a single star, there are more orbital interactions including nodal resonances and mean motion resonances with the binary, mean motion resonances and tilt oscillation between planets and some ZKL--like mechanism in multiple CBP systems. Our four-body simulations show that large unstable regions dominate each stability map as the two planets become more misaligned to the binary orbital plane. For example, the cases plotted in Figures~\ref{fig:mapa5}, \ref{fig:mapa10}, and \ref{fig:mapa20}, should all be stable to single planet interactions with the binary \citep{Chen20201}. These cases should also be stable against planet-planet interactions involving a single star, except at the closest planet separations, as seen in Figure~\ref{fig:single}. Yet, the systems are unstable over much of the plotted parameter ranges. This broader range of instability is due to four-body effects involving the interplay between these two interactions.
The final outcome for most of the unstable simulations is that one planet is ejected, rather than two.

What appears to be a modified version of the standard ZKL effect plays an important role in the stability maps when the planets are far from the binary. There are two ways for planets to  be stabilised when they are in the ZKL region. First, if the phase angles of the two planets are locked to each other their mutual inclination is too small to trigger the ZKL mechanism.  Secondly, libration of planet orbits with the binary reduces their mutual inclinations and suppresses the ZKL effect when $e_{\rm b}$ is large. 

\begin{figure*}
    \centering
    \includegraphics[width=8.7cm]{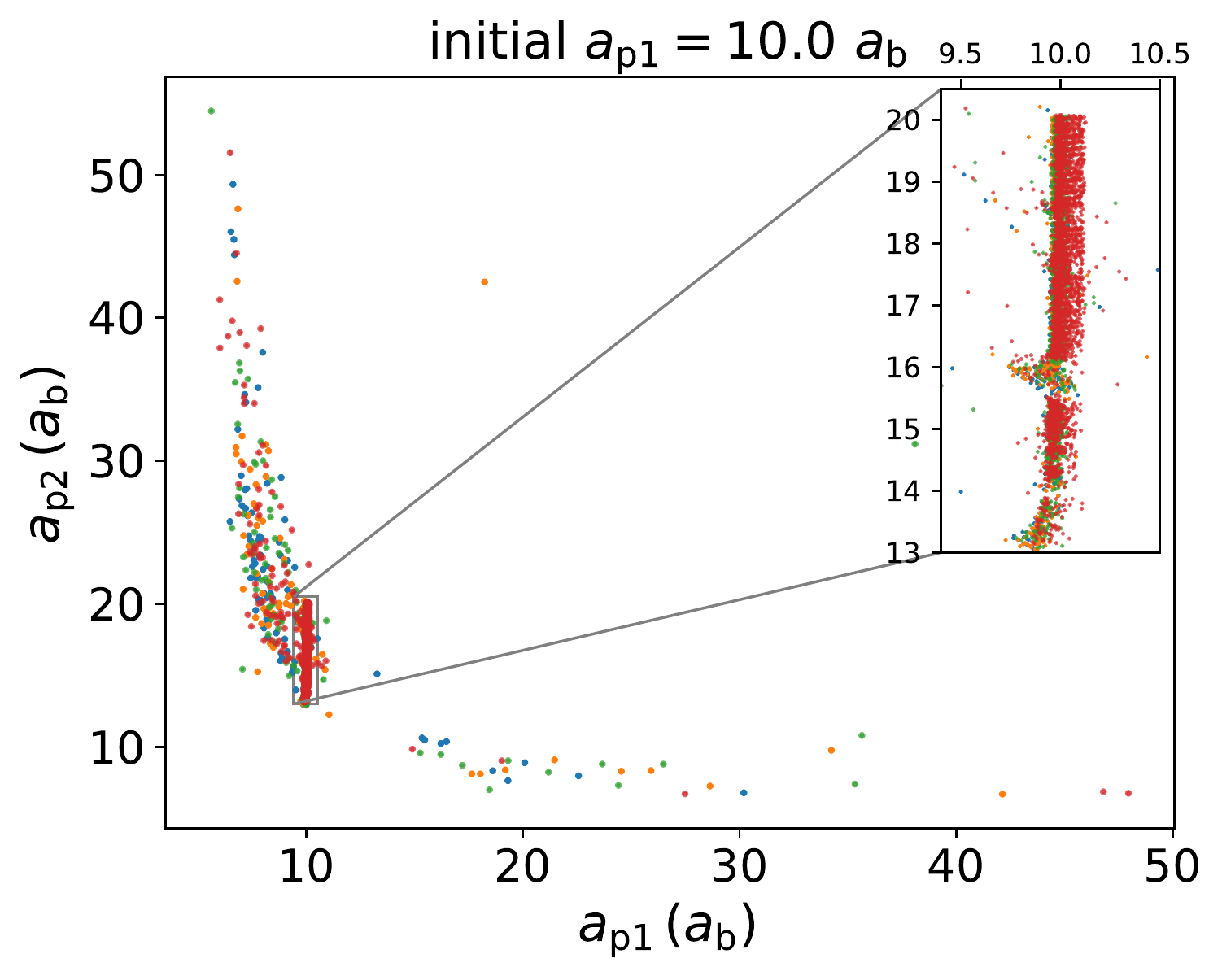}
    \includegraphics[width=8.7cm]{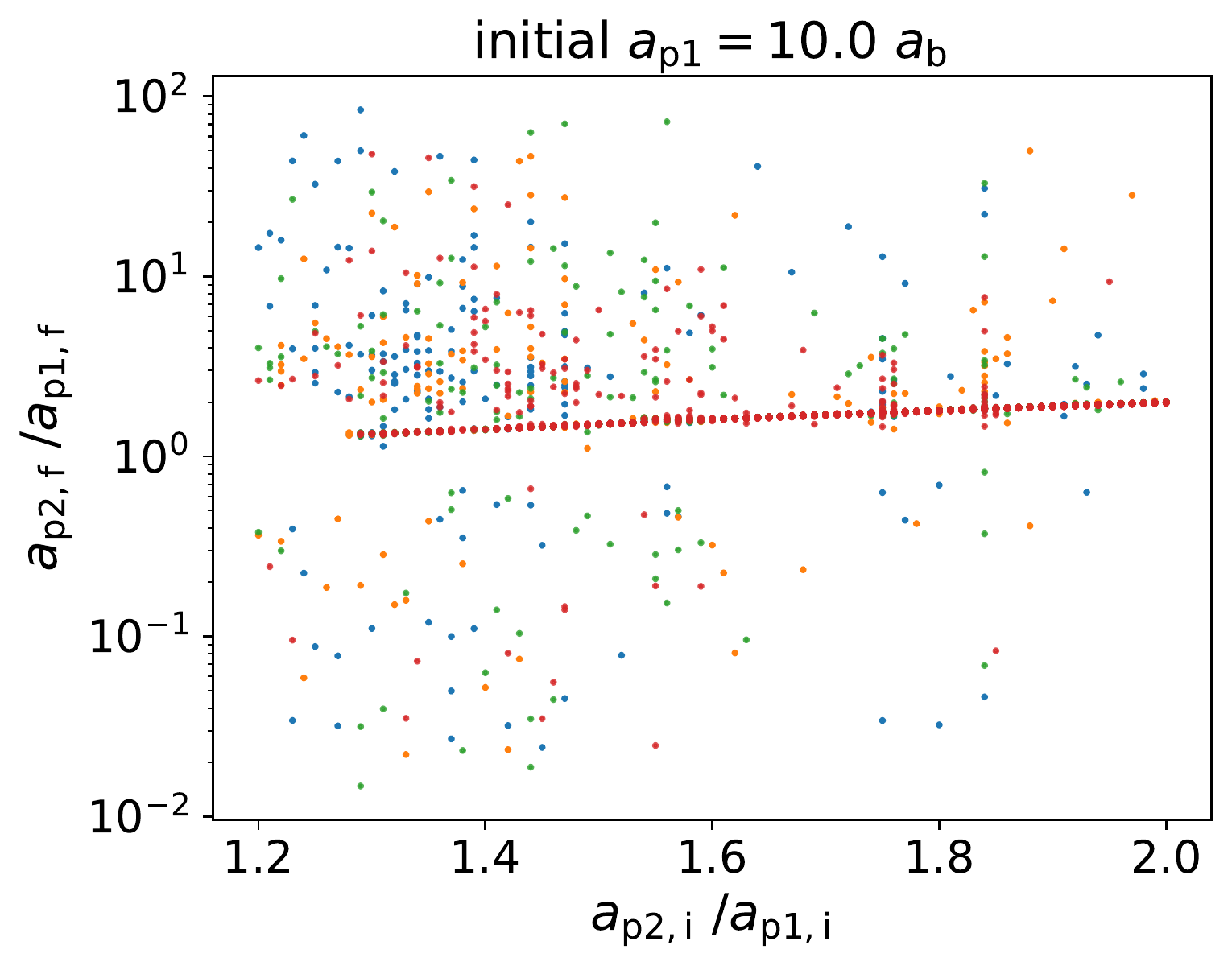}
    \caption{Left panel: the final distribution of $a_{\rm p1}$ versus $a_{\rm p2}$ for two planets stable cases of model Y1 $\sim$ Y4. The blue, yellow, green and red dots represent  cases of $e_{\rm b}$  = 0.0, 0.2, 0.5 and 0.8, respectively. Right panel: the initial ratio of $a_{\rm p2}$ / $a_{\rm p1}$ versus the final ratio of $a_{\rm p2}$ / $a_{\rm p1}$. }
    \label{fig:aaa10}
\end{figure*}

The majority of the outcomes in our stability maps are two stable planets cases. If both planets are stable at the end of the simulation, the final distribution of the orbital properties show that $e_{\rm p}$ of both planets have non zero but < 0.1 mean values while the mean value of $a_{\rm p}$ is in the range 0.5 $\sim$ 1.5 times final $a_{\rm p}$ / initial $a_{\rm p}$. Thus, the final $a_{\rm p1}$ and $a_{\rm p2}$ on the stability map may not represent their final locations although most of cases are very close to their initial locations. In Fig.~\ref{fig:aaa10}, the left panel shows that the final distribution of $a_{\rm p1}$ versus $a_{\rm p2}$ of two planet stable cases in model Y1 $\sim$ Y4. The blue, yellow, green and red dots represent $e_{\rm b}$ = 0, 0.2, 0.5 and 0.8, respectively. (The red dots look like being most dominant because they overlap other dots with different colours.) The final $a_{\rm p}$ in the most of two planets stable cases are about $\leq$ 1$\%$ different from initial $a_{\rm p}$ except in resonance and innermost regions where are less two planets stable cases. A clump of inner planets are scattered inward while the outer planet are scattered outward and vice versa. On the right panel, we plot the ratio of final semi-major axes $a_{\rm p2, f}$ / $a_{\rm p1, f}$ versus the ratio of initial semi-major axes $a_{\rm p2, i}$ / $a_{\rm p1, i}$ which shows that inner than $a_{\rm p1, i}$ = 13$a_{\rm b}$, some two planets cases can be stable but they are much further than their original semi-major axes. The closest inner planet in model X1 $\sim$ X4 at the end of the simulation can be close to 3$\,a_{\rm b}$ and those of in model Y1 $\sim$ Y4 and model Z1 $\sim$ Z4 are about 5$\,a_{\rm b}$. Moreover, these distributions are not sensitive to $e_{\rm b}$ except for models X1 $\sim$ X4 because planets could be very close to the binary. As a result, we predict that two CBPs in a binary system may not have highly eccentric orbits unless they underwent large migration and had close encounters with a binary during the early stages of planet formation. 

The single survival planet cases make up the majority of the unstable cases. After the ejection of one planet, the final distributions of the orbital properties of the surviving planets can help us to understand and predict future observations for a binary with a single planet because $e_{\rm p}$ of two CBPs are excited during the evolution. However, the distributions of $a_{\rm pf}$ show that the final $a_{\rm p}$ could be several times larger than the initial $a_{\rm p}$ unlike the two stable planets cases. It may be hard to detect those distant and misaligned CBPs in binary systems. The final orbital distributions are also not very sensitive to $e_{\rm b}$. There are not significant differences between the inner planet surviving cases and outer planet surviving cases and the mean value of $e_{\rm p}$ is about 0.5. As the result, a CBP which has a significant eccentricity and an inclined orbit may have undergone planet-planet interactions after the gas disc was dissipated.

The mass of the planet also plays a role in the orbital dynamics and stability. A high mass planet can have a strong interaction with the other planet due to the high angular momentum exchange. On the contrary, a small mass planet has a smaller effect on the other planet. Furthermore, the critical inclination for the planet to be in a polar librating orbit and the stationary inclination of the planet depend on the binary eccentricity and angular momentum ratio of the planet to the binary \citep{MartinandLubow2019}. We ran model Z3 again but changed two CBPs to have a smaller  mass of $0.0001 \,m_{\rm b}$. The dynamics of the planets were dominated by the binary with little perturbation from the other planet.   Additionally, during the gas disc phase, small planets may undergo type-I migration. The high type~1 migration speed is linearly proportional to the planet mass \citep{Tanaka2002, LubowIda2010}. Consequently, these planets may not stay in the outer disc for a long time but they may migrate inwards until meeting a steep density gradient at the inner edge of the circumbinary disc \citep{Pierens2008,Pierens2013,Thun2016,Penzlin2021}. Formation of terrestrial planets from a misaligned disc of planetestimals after the gas disc has dissipated may also be difficult, unless the disc is close to coplanar  or polar alignment \citep{Childs2021,Childs2021b, Childs2022}. A highly misaligned disc of planetesimals around a binary may be largely ejected \citep{Childs2022b}. Therefore, multiple low mass CBP systems may be not applicable for our study. A comprehensive study of the orbital instabilities of two CBPs which are very close to the binary has been done by \citet{Sutherland2019}.

In this study, we do not allow for collisions or resolve the formation of S-type planets (planets that orbit around one star of a binary) in our simulations. Further, we only consider equal mass binaries. CBPs that are captured by one star of a binary may reveal a formation mechanism for S-type planets \citep{Dvorak1986}. Current observations show that there are no S-type planets found in binary systems with $a_{\rm b} < 5.2 \,\rm au$. However, since the Kepler telescope has detected more than 3000 eclipsing binaries with $T_{\rm b} < 3000 \,\rm days$ \citep{Kirk2016}, this implies that S-type planets could be found in those short period binary systems. Simulations in \cite{Gong2018} show that the maximum capture probability due to coplanar planet-planet scattering is about 10 \% for small binary mass ratio $q_{\rm b}$ and $e_{\rm b}$  while it is only 1 \% for equal mass binaries.  Therefore, it is worth considering the effect of $q_{\rm b}$ in future work to investigate the orbital stability of the multi-circumbinary planet systems with different $q_{\rm b}$. Furthermore, our model has the potential to calculate the capture probability of CBPs with misaligned orbits and it could provide a comprehensive understanding of the formation of S-type planets in short period binaries. 

During planet-planet interactions, an ejection of one planet from the system requires the ejected planet to be accelerated to a velocity above the escape speed from the binary $v_{\rm esc, b} = \sqrt{2Gm_{\rm b}/a_{\rm p}}$  For the bound planet to accelerate the ejected planet to this velocity roughly requires the planets' escape velocity, $v_{\rm esc, p} = \sqrt{2Gm_{\rm p}/R_{\rm p}}$ (where $R_{\rm p}$ is the radius of the planet), to be greater than the escape speed from the binary. The criterion $v_{\rm esc, p} > v_{\rm esc, b}$ implies that the orbital radius of the planet around the binary must be $a > (m_{\rm b}/m_{\rm p})R_{\rm p}$. For a solar mass binary and Jupiter-like planet, this results in $a > 0.5$au. We therefore expect that most of the unstable cases in these types of systems will lead to ejections rather than direct collisions of the planets. However, for short period binaries with planets orbiting at radii of order an au, collisions may be common and we would expect interesting dynamics and observational signatures — including accretion of planetary debris on to the central binary — to occur.

\section{Implications for observations}

The origin of free-floating planets is still not clear because there is, as yet, no statistical analysis with a large homogeneous sample \citep{Miret-Roig2021}. However, our results  show that misaligned binaries can be  efficient drivers for the formation of  free-floating planets, especially for Jovian size planets. The conditions required for a stable two planet system are much stricter around a binary than around a single star system. Planet-planet interactions around a binary system can lead to planet-binary or planet-planet close encounters and planet ejection. Moreover, by the end of our simulations, some planets have extremely eccentric orbits with large $a_{\rm p}$. These may also contribute to the population of the free-floating planets since they could become unstable due to the small disturbance and our integration time is limited. Long-term dynamical effects may not be seen since the orbital period of the planet is too long.

There are currently 12 observed unbound planetary objects with a mass in the range 3 to 15 Jupiter masses. They were all observed through gravitational microlensing and their abundance is estimated to be about 1.8 times as common as main-sequence stars \citep{Sumi2011}.  Planet-planet scattering and post-formation evolution are not enough to explain the free-floating planet population. Indeed,  scattering of Jupiter-mass planets in multi-planetary systems around a single star is not efficient (see Fig.~\ref{fig:single}). It has been suggested that other effects may have to take into account such as planetary stripping in stellar clusters and post-main-sequence ejection \citep{Boss2006, Veras2012, Pfyffer2015, Barclay2017}.  Predictions for the microlensing event rate of free-floating planets from the core accretion theory are smaller than the estimated value from observations \citep{Sumi2011,Ma2016}. The effect of fragment-fragment interactions in a self-gravitating disc can also contribute to producing massive free-floating planets \citep{Forgan2018}. However, we suggest that formation of planets in a misaligned disc around a binary may significantly increase the number of free-floating planets. The Nancy Grace Roman Space Telescope (Roman) survey will look for free-floating planets in the Galactic bulge in an upcoming observation season \citep{Penny2019, Johnson2020} and the Roman mission has the potential to search free-floating planets in the Magellanic Clouds \citep{Sajadian2021}.  

Apart from TESS and {\it PlAnetary Transits and Oscillations of stars} (PLATO), the Kepler telescope has also contributed to finding CBPs. It is much easier to confirm a CBP if a transit cannot be mocked by an eclipsing binary \citep{Doyle2011}. There are two systems KIC 07177553 and KIC 7821010 that may have a planetary-mass object orbiting around in a misaligned orbit around a binary \citep{Borkovits2016}. We consider the first two systems since there is enough orbital information. KIC 07177553 has a total mass of $1.9\,\rm M_{\odot}$, a binary mass fraction of $f_{\rm b}=0.486$, a binary eccentricity of $e_{\rm b} =0.39$ and the planetary object has a mass of $5.24 \,M_{\rm J}$, with orbital separation $a = 9.52 \,a_{\rm b}$, inclination to the binary orbit $i= 26^\circ$ and $\phi = 26^\circ$. KIC 7821010 has a total mass of  $2.52 \,\rm M_{\odot}$, a binary mass fraction of $f_{\rm b}=0.488$, a binary eccentricity of $e_{\rm b}=0.68$ and the planetary object has a mass of $2.56 \,\rm M_{\rm J}$, orbital semi-major axis $a= 11.88 \,a_{\rm b}$, inclination to the binary orbital plane $i$ = $25^\circ$ and a longitude of ascending node of $\Omega = -19^\circ$. With similar analysis to that presented in \citet{Chen20192, Chen20201}, we find that the two planetary objects are both in prograde circulating orbits. Recently,  using ETVs, two CBPs were found  around an eclipsing binary RR Cae which is comprised of a white dwarf and  M-type dwarf  \citep{Rattanamala2021}. We expect to find more inclined or polar circumbinary multi-planetary systems with future observations by TESS or PLATO as a result of the well-developed ETV tools \citep{Zhang2019}. 

\section*{Data Availability}
    The simulations in this paper can be reproduced by using
    the REBOUND code  (Astrophysics Source Code
    Library identifier ascl.net/1110.016). The data underlying 
    this article will be shared on reasonable request to the corresponding author.

\section*{Acknowledgements}
Computer support was provided by UNLV's National Supercomputing Center. C.C. acknowledges support from a UNLV graduate assistantship. CC and CJN acknowledge support from the Science and Technology Facilities Council (grant number ST/W000857/1). CJN acknowledges support from the Leverhulme Trust (grant number RPG-2021-380). We acknowledge support from NASA through grants NNX17AB96G and 80NSSC19K0443. We thank the anonymous reviewer for his/her careful reading of our manuscript and gives us many insightful comments and suggestions. Simulations in this paper made use of the REBOUND code which can be downloaded freely at http://github.com/hannorein/rebound.



\bibliographystyle{mnras}
\bibliography{main} % if your bibtex file is called example.bib

\bsp	% typesetting comment
\label{lastpage}
\end{document}